\documentclass[useAMS,usenatbib]{mn2e}

\usepackage{float}
\usepackage{graphicx}
\usepackage{url}
\usepackage[english]{babel}
\usepackage{morefloats}

\title[Saturn Irregular Satellites - Origin]{A new perspective on the irregular satellites of Saturn - II\\Dynamical and physical origin}
\author[Turrini et al.]{D. Turrini$^{1,2}$\thanks{E-mail: diego.turrini@ifsi-roma.inaf.it}, F. Marzari$^{2}$, F. Tosi$^{3}$\\
$^{1}$Center of Studies and Activities for Space ``G. Colombo'', University of Padova, Via Venezia 15, 35131, Padova, Italy\\
$^{2}$Physics Department, University of Padova, Via Marzolo 8, 35131, Padova, Italy\\
$^{3}$Institute for Interplanetary Space Physics, INAF, Via del Fosso del Cavaliere 100, 00133, Roma, Italy\\}
\begin{document}

\date{Accepted XXX. Received XXX; in original form XXX}

\pagerange{\pageref{firstpage}--\pageref{lastpage}} \pubyear{2007}

\maketitle

\label{firstpage}

\begin{abstract}
The origin of the irregular satellites of the giant planets has been long debated since their discovery. Their dynamical features argue against an in-situ formation suggesting they are captured bodies, yet there is no global consensus on the physical process at the basis of their capture. In this paper we explore the collisional capture scenario, where the actual satellites originated from impacts occurred within Saturn's influence sphere. By modeling the inverse capture problem, we estimated the families of orbits of the possible parent bodies and the specific impulse needed for their capture. The orbits of these putative parent bodies are compared to those of the minor bodies of the outer Solar System to outline their possible region of formation. Finally, we tested the collisional capture hypothesis on Phoebe by taking advantage of the data supplied by Cassini on its major crater, Jason. Our results presented a realistic range of solutions matching the observational and dynamical data.
\end{abstract}

\begin{keywords}
planets and satellites: formation, Saturn, irregular satellites - methods: numerical, N--Body simulations - celestial mechanics
\end{keywords}

\section{Introduction}\label{introduction}

In our previous paper (Turrini et al., accepted, hereafter Paper I) on the subject of Saturn's irregular satellites we explored the dynamical and collisional structure of the system to shed new light on its past history. The results we obtained strongly suggested that the system of irregular satellites  of Saturn went subject to a strong dynamical, and probably also collisional, sculpting during its lifetime. The actual structure of the system appeared to be very likely less representative than previously thought of the original, post-capture one. Nevertheless, through our results we have been able to draw a coherent picture of the possible past evolution of the system.\\
Before proceeding further, we would like to briefly summarise the main results of our previous work:
\begin{itemize}
\item we found confirmation for the resonant motion in the Kozai regime for Ijiraq and, partially, Kiviuq;
\item our numerical simulations showed strong evidences of the presence of chaos in the system, with about two thirds of the satellites being affected to some extent;
\item we verified that the influence of the Great Inequality between Saturn and Jupiter plays a central role in shaping the dynamical behaviours of the irregular satellites;
\item we found that the influence of the two outermost major satellites (Titan and Iapetus) alters the evolution of the system, having stabilising effects on some satellites and destabilising ones on others;
\item we verified that the present orbital structure is long--lived against collisions but there are indications that in the past a more intense collisional activity could have taken place;
\item we interpreted the absence of prograde irregular satellites in the region comprised between $11.22 \times 10^{6}$ km and $14.96 \times 10^{6}$ km as a by-product of the sweeping effect of Phoebe;
\item similarly, we suggest that the absence of retrograde satellites in the same radial region could be due to the same mechanism, even if we could not rule out the possibility of this being a primordial heritage;
\item by applying the HCM algorithm \citep{zap90,zap94} we found two candidate families between the prograde and six between the retrograde satellites, with some of the families merging into bigger groups we called clusters.
\end{itemize}
The framework of our first article was our project to study the origin of the irregular satellites, using Saturn system as our case study to take advantage of the unprecedented data supplied by the Cassini mission during its only flyby of Phoebe. In particular, our interest concentrated on investigating the viability and the consequences of the collisional capture process, originally proposed by \cite{col71}, which could supply a natural way to explain the formation of the systems of irregular satellites of both gas and ice giant planets.
With the deeper insight of the dynamical and collisional nature of the system of irregular satellites of Saturn obtained through our previous results, we investigated the collisional capture process with the aims of constraining the characteristics of the impacts and of searching for information on the possible pre-capture orbits.\\
The work to be presented is organised as follows:
\begin{itemize}
\item in section \ref{capture} we will describe the collisional capture scenario and the dynamical strategy we devised to explore its implications;
\item in section \ref{comparison} we will compare the pre--capture orbits obtained through the numerical code we developed with the ones of the minor bodies in the outer Solar System;
\item in section \ref{phoebe} we will search for evidences supporting the collisional capture hypothesis in the data supplied by Cassini;
\item finally, in section \ref{conclusion} we will put together all our results to draw a global picture of the origin of Saturn's system of irregular satellites.
\end{itemize}

\section{The collisional capture scenario}\label{capture}

Our interest in studying the capture mechanisms of irregular satellites concentrated on the collisional scenario originally proposed by \cite{col71}, which seems to be the only one capable to explain in a natural way the existence of the systems of irregular satellites of both gas and ice giants.\\
The ideal, ground--based strategy to investigate this capture scenario would be to identify the possible collisional families existing in the system through the comparative study of their mean orbital elements and their surface compositions. In principle, the masses, the orbits and the composition of the parent bodies, once estimated, could make it possible to identify their nature and their region of formation. Even more ideal it would be to search for features proving the occurrence of relevant or disruptive impacts on the surface of the satellites through space--based observations from probes performing close fly--bys.\\
The results we presented in our previous paper showed that these ideal conditions are difficult to meet in the case of the Saturn system. The collisional history of its irregular satellites appears to have been eventful and the actual structure of the system is probably not representative of the original, post--capture one. Moreover, the combined effects of secondary collisions and of the perturbed, chaotic dynamical evolution of some of the satellites could have contributed in mixing and altering the possible families generated by catastrophically disruptive events, thus increasing the difficulty of the task of identifying them.\\ 
In addition to these problems, the distance and the small size of the irregular satellites make it difficult to observe them and obtain detailed spectral information with Earth--based telescopes. In general, the only observational data that could be gathered are the information--poor colour indexes. Even for the biggest and most luminous irregular satellites, the spectral range achievable is quite limited and it is therefore difficult to trace back their compositions.\\ 
In Saturn's case, however, one of the ideal conditions for the study of the collisional capture scenario can be met. Thanks to the Cassini spacecraft, which performed the first and to now only observations of an irregular satellite during a close fly--by, detailed information on the physical and compositional nature of Phoebe are available. This is the reason behind our choice of the Saturn system as our case study.\\

\subsection{The inverse capture problem: dynamical model and numerical code}

Since realistic simulations of the collisional capture process are to now an unreachable goal due to the limited data available, the arbitrary initial conditions and the degeneracy of the solutions, we devised the following strategy to investigate the problem. 
Instead of exploring directly the collisional capture, we concentrated on the inverse problem (i.e. the orbital unbinding from the host planet) taking advantage of the time reversibility of the equations of motion. The dynamical model we devised employs the specific impulse needed to obtain the passage from a bound planetocentric orbit to a bound heliocentric orbit as its only parameter and is designed to search for the possible related solutions. Our choice to employ the specific impulse as the free parameter aimed at minimising the dependence of the solutions from the characteristics of the involved bodies and to obtain information as general as possible. The assumption underlying the model is that the capture of an irregular satellite should have been due to a single, predominant impact event which we could model through the application of a delta--shaped specific impulse (i.e. an instantaneous change in velocity).\\
We numerically implemented the model in the code MSSCC (\textit{Modelling Software for Satellite Collisional Capture}), based on the patched conics approximation, which solving algorithm addresses the inverse problem in the following way:
\begin{itemize}
\item given the initial planetocentric orbital elements of the satellite, its ($\vec{r}$,$\vec{v}$) planetocentric vectors are computed;
\item a spherical sampling of vectors $d\vec{v_{i}}$ is created using the fixed modulus $dv$ and the angular sampling steps $\Delta\theta$ and $\Delta\varphi$ specified in input by the user;
\item the specific impulse vectors $d\vec{v_{i}}$ are summed to the $\vec{v}$ vector and the new pairs of conjugate variables ($\vec{r_{i}}$,$\vec{v_{i}}^{'}$) are computed with $\vec{r_{i}}=\vec{r}$;
\item converting the ($\vec{r_{i}}$,$\vec{v_{i}}^{'}$) pairs to orbital elements, the orbits which are hyperbolic in the planetocentric frame are kept while the others are discarded; 
\item the selected orbits are reverted to the heliocentric frame and only the elliptic, prograde ones are saved as candidate primordial orbits for the parent bodies.
\end{itemize}
The condition on the inclination of the candidate orbits (i.e. they must be prograde) is imposed to assure the physical correctness of the solutions found, since the formation of retrograde bodies from the Solar Nebula is not plausible.\\
If specified by the user, the code can loop the algorithm while varying the orbital position of both the satellite around the planet and the planet around the Sun to avoid selection effects due to the choice of particular initial conditions. The code also implements the possibility to filter the candidate orbits over a selected $e_{max}$ value of the eccentricity, in order to control the range of physical conditions under which the capture event took place.\\
An open issue with this approach is the physical meaning of the specific impulse $d\vec{v}$ used by the algorithm. To the first order and for ideally, fully inelastic collisions the interpretation is straightforward, since it represents the change in velocity due to the collision of the parent body with the projectile. The real case is however complicated by the fact that impacts always involve some level of energy dissipation due to the generation of heat, the formation of craters and fracture lines in the target body and the excavation of fragments. Moreover, things are worse in case of catastrophically disruptive events. While for the ejection of fragments the total excavated mass is usually small in comparison to the mass of the target and the contribution to the change in momentum is limited, in case of catastrophic disruption the mass of the collisional shards could be relevant and the ejection speeds, estimated from laboratory experiments and numerical simulations with hydrocodes to be of the order of $100$ m/s \citep{ben99}, could represent a significant fraction of the specific impulse imposed by the collision. While this problem in general has to be dealt with on a case by case basis, we will treat it in major detail for Phoebe in section \ref{phoebe}.

\subsection{Settings and results of the dynamical analysis}

\begin{figure*}
\begin{center}
\includegraphics[width=16cm]{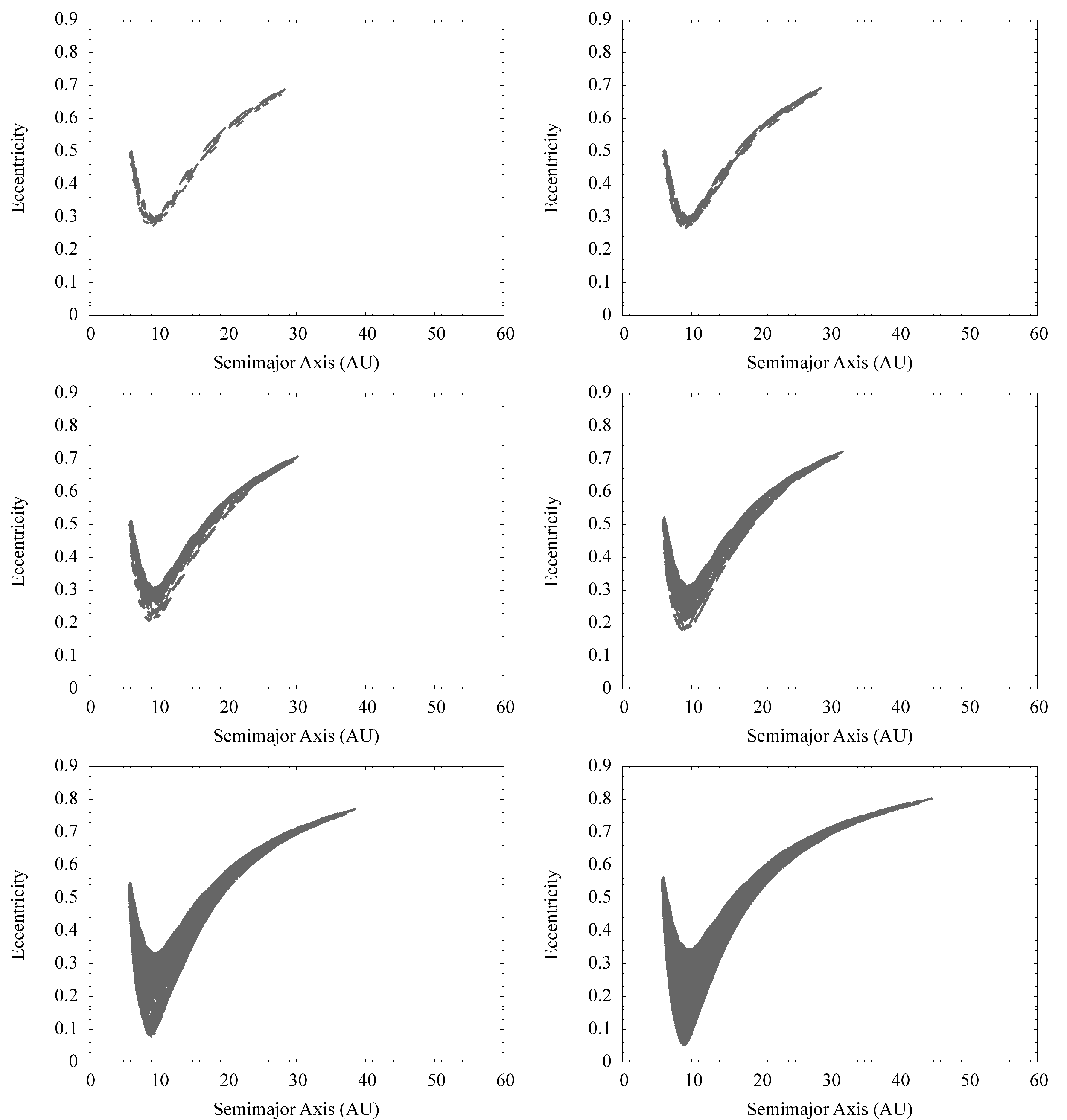}
\end{center}
\caption{Evolution of the solutions to the inverse capture problem obtained for Albiorix for increasing values of the specific impulse applied and projected in the heliocentric $a-e$ plane. From top left to bottom right, the values assumed by the specific impulse are $445,\,460,\,520,\,580,\,760,\,880$ m/s. Distances are expressed in AU.}
\label{aeplot-Albiorix}
\end{figure*}

\begin{figure*}
\begin{center}
\includegraphics[width=16cm]{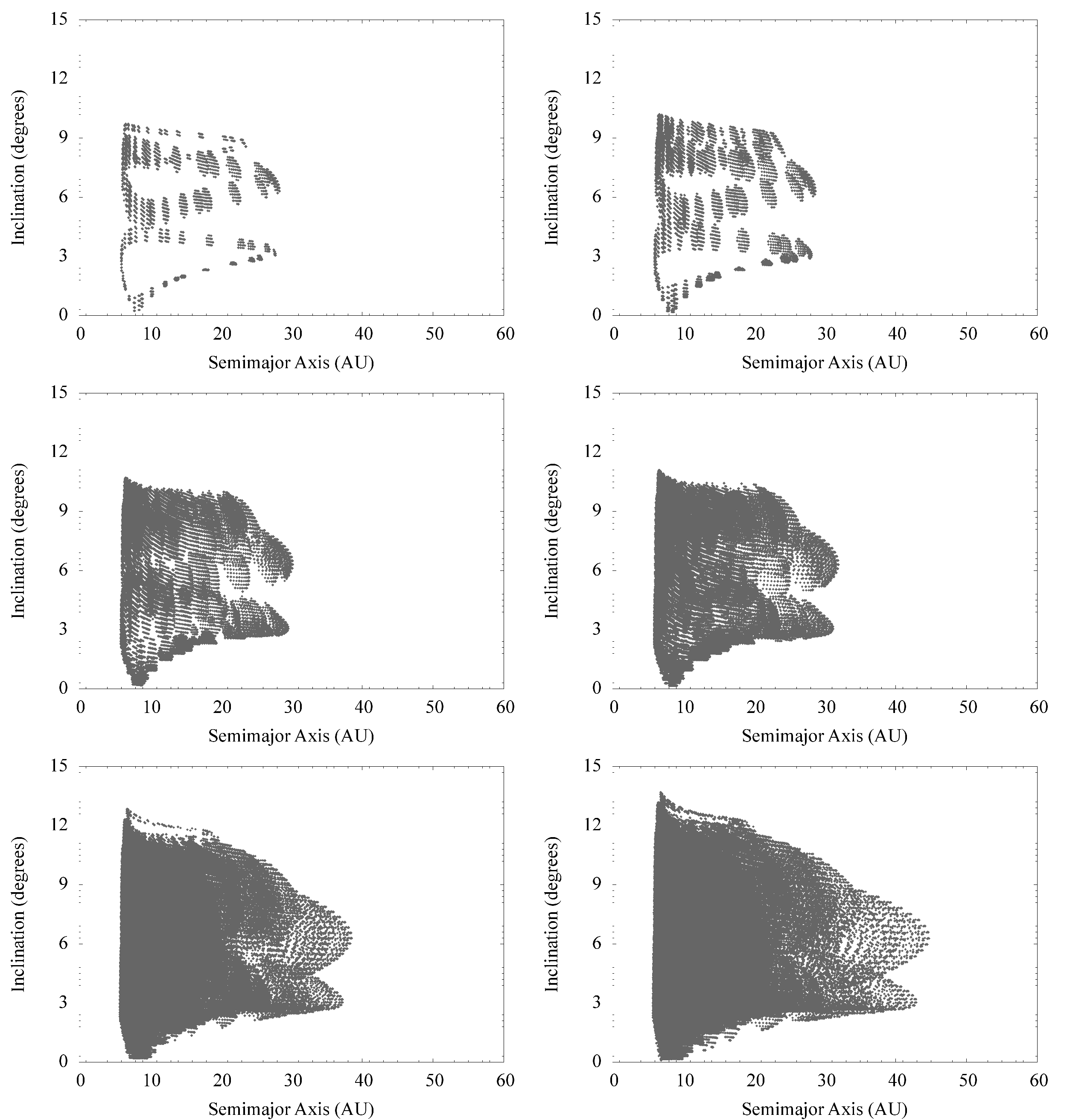}
\end{center}
\caption{Evolution of the solutions to the inverse capture problem obtained for Albiorix for increasing values of the specific impulse applied and projected in the heliocentric $a-i$ plane. From top left to bottom right, the values assumed by the specific impulse are $445,\,460,\,520,\,580,\,760,\,880$ m/s. Distances are expressed in AU while angles are expressed in degrees.}
\label{aiplot-Albiorix}
\end{figure*}

\begin{figure*}
\begin{center}
\includegraphics[width=16cm]{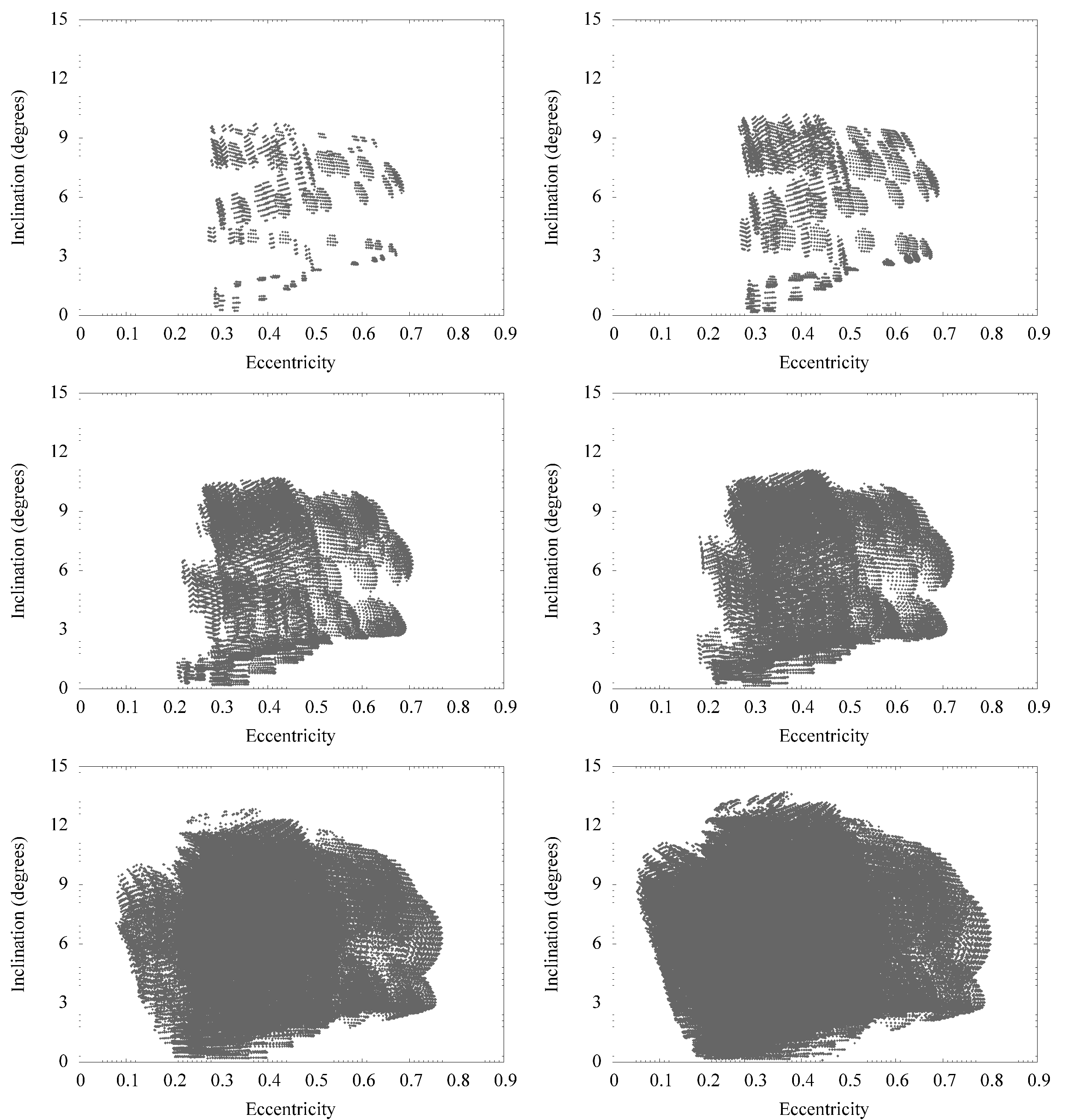}
\end{center}
\caption{Evolution of the solutions to the inverse capture problem obtained for Albiorix for increasing values of the specific impulse applied and projected in the heliocentric $e-i$ plane. From top left to bottom right, the values assumed by the specific impulse are $445,\,460,\,520,\,580,\,760,\,880$ m/s. Angles are expressed in degrees.}
\label{eiplot-Albiorix}
\end{figure*}

\begin{figure*}
\begin{center}
\includegraphics[width=16cm]{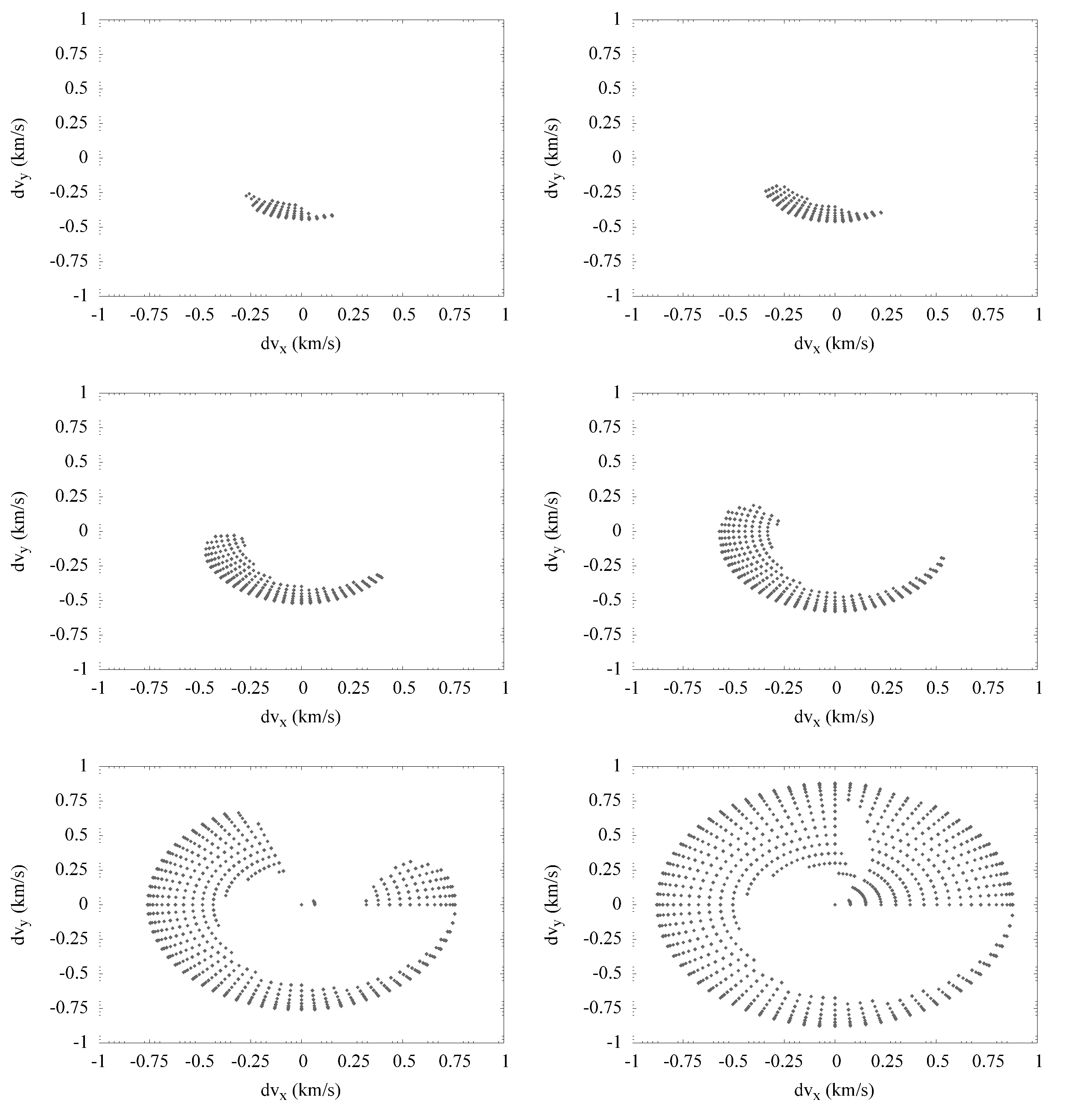}
\end{center}
\caption{Specific impulse applied to Albiorix to solve the inverse capture problem projected in the $v_{x}-v_{y}$ plane. From top left to bottom right, the values assumed by the specific impulse are $445,\,460,\,520,\,580,\,760,\,880$ m/s. Velocities in the graphs are expressed in km/s.}
\label{dvplot-Albiorix}
\end{figure*}

\begin{figure*}
\begin{center}
\includegraphics[width=16cm]{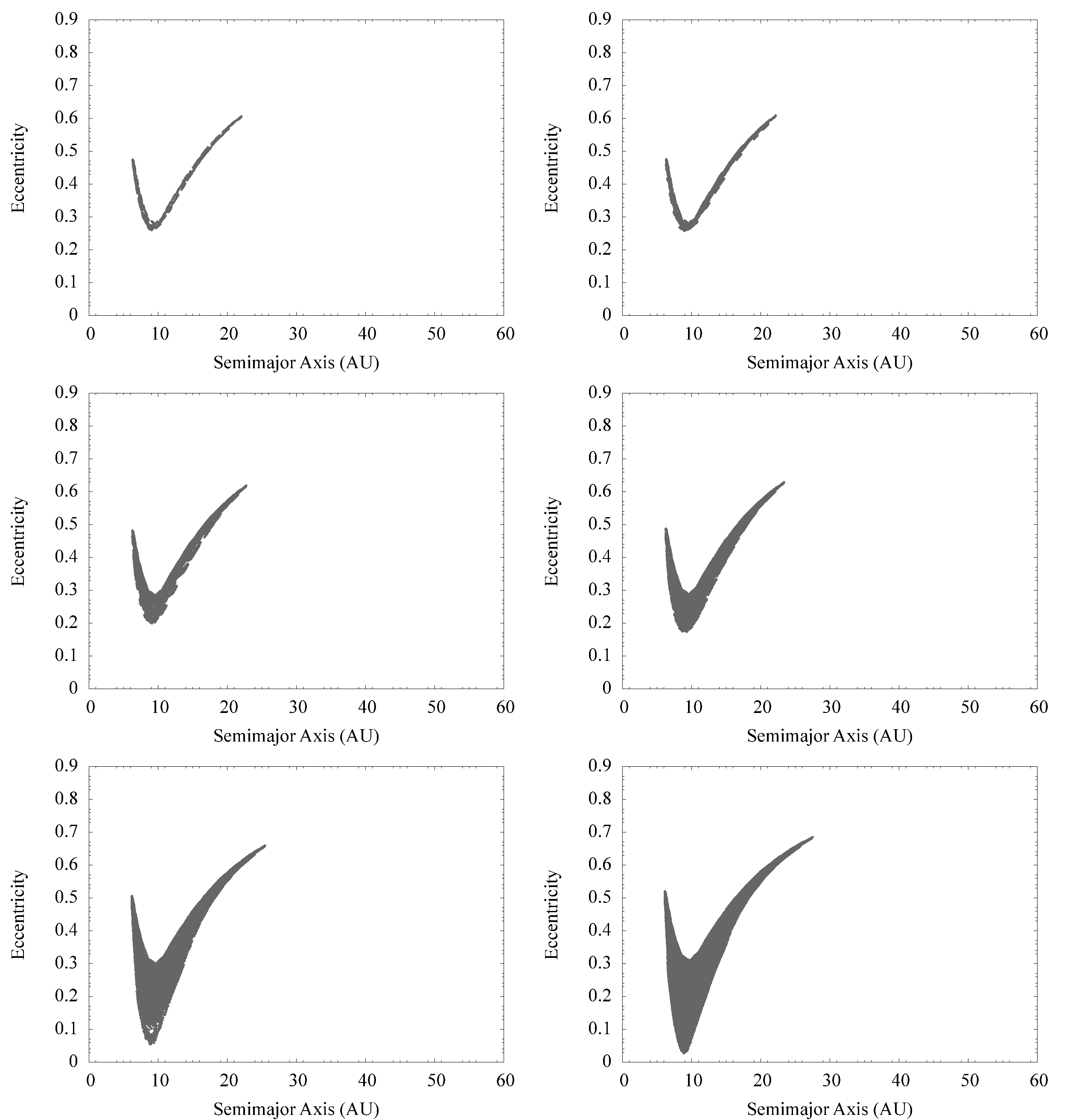}
\end{center}
\caption{Evolution of the solutions to the inverse capture problem obtained for Siarnaq for increasing values of the specific impulse applied and projected in the heliocentric $a-e$ plane. From top left to bottom right, the values assumed by the specific impulse are $490,\,500,\,540,\,580,\,700,\,800$ m/s. Distances are expressed in AU.}
\label{aeplot-Siarnaq}
\end{figure*}

\begin{figure*}
\begin{center}
\includegraphics[width=16cm]{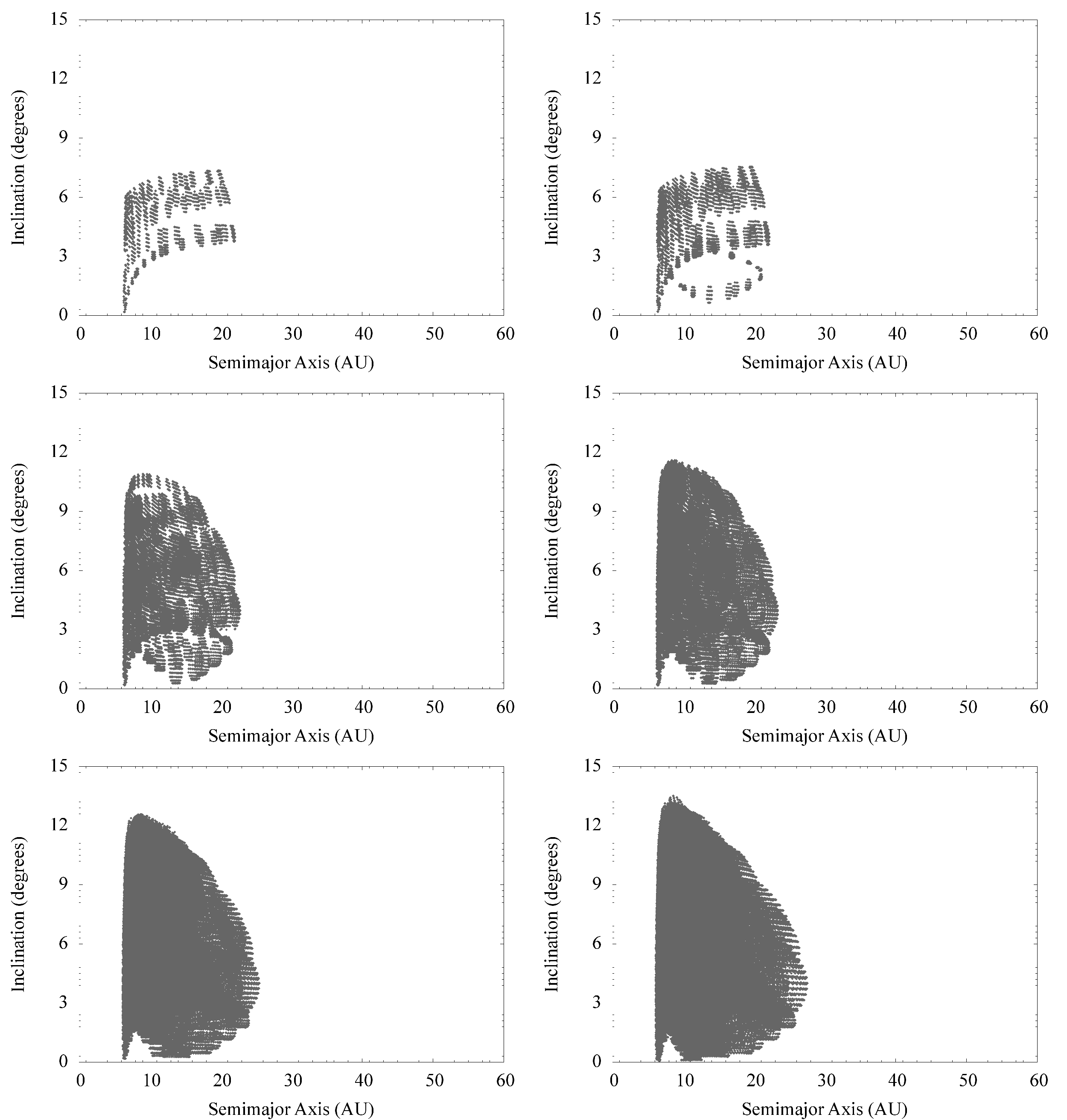}
\end{center}
\caption{Evolution of the solutions to the inverse capture problem obtained for Siarnaq for increasing values of the specific impulse applied and projected in the heliocentric $a-i$ plane. From top left to bottom right, the values assumed by the specific impulse are $490,\,500,\,540,\,580,\,700,\,800$ m/s. Distances are expressed in AU while angles are expressed in degrees.}
\label{aiplot-Siarnaq}
\end{figure*}

\begin{figure*}
\begin{center}
\includegraphics[width=16cm]{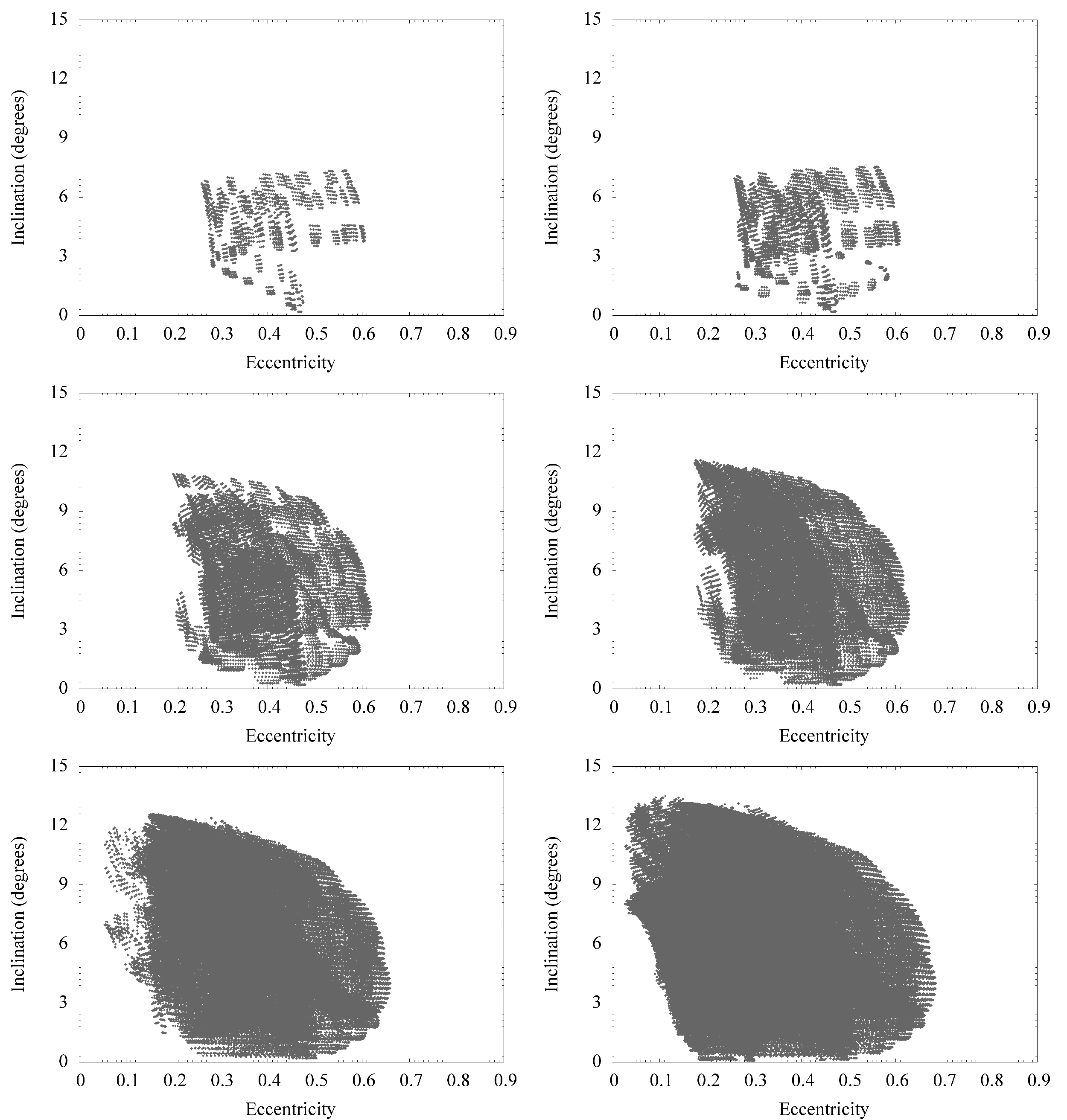}
\end{center}
\caption{Evolution of the solutions to the inverse capture problem obtained for Siarnaq for increasing values of the specific impulse applied and projected in the heliocentric $e-i$ plane. From top left to bottom right, the values assumed by the specific impulse are $490,\,500,\,540,\,580,\,700,\,800$ m/s. Angles are expressed in degrees.}
\label{eiplot-Siarnaq}
\end{figure*}

\begin{figure*}
\begin{center}
\includegraphics[width=16cm]{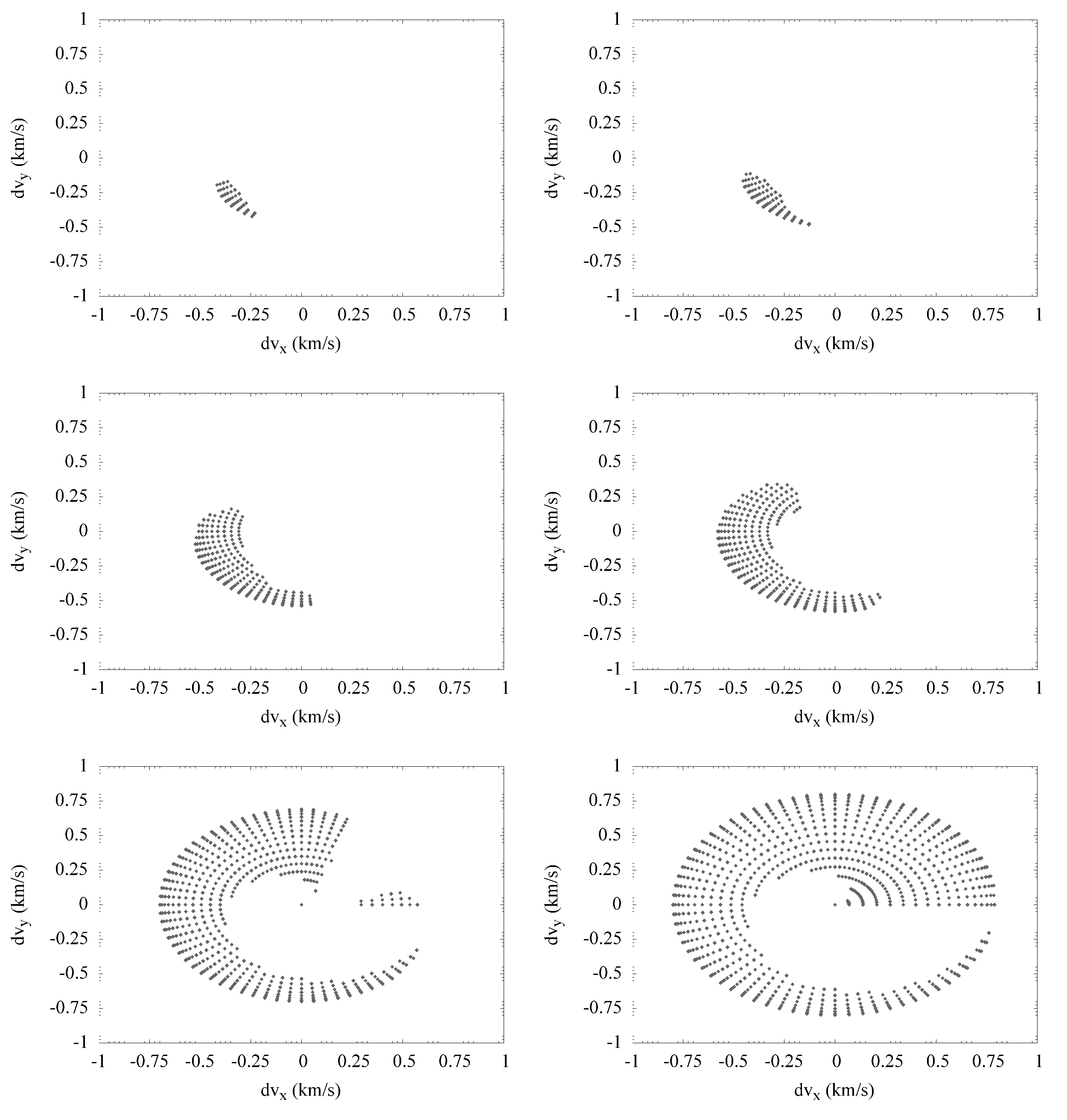}
\end{center}
\caption{Specific impulse applied to Siarnaq to solve the inverse capture problem projected in the $v_{x}-v_{y}$ plane. From top left to bottom right, the values assumed by the specific impulse are $490,\,500,\,540,\,580,\,700,\,800$ m/s. Velocities in the graphs are expressed in km/s.}
\label{dvplot-Siarnaq}
\end{figure*}

\begin{figure*}
\begin{center}
\includegraphics[width=16cm]{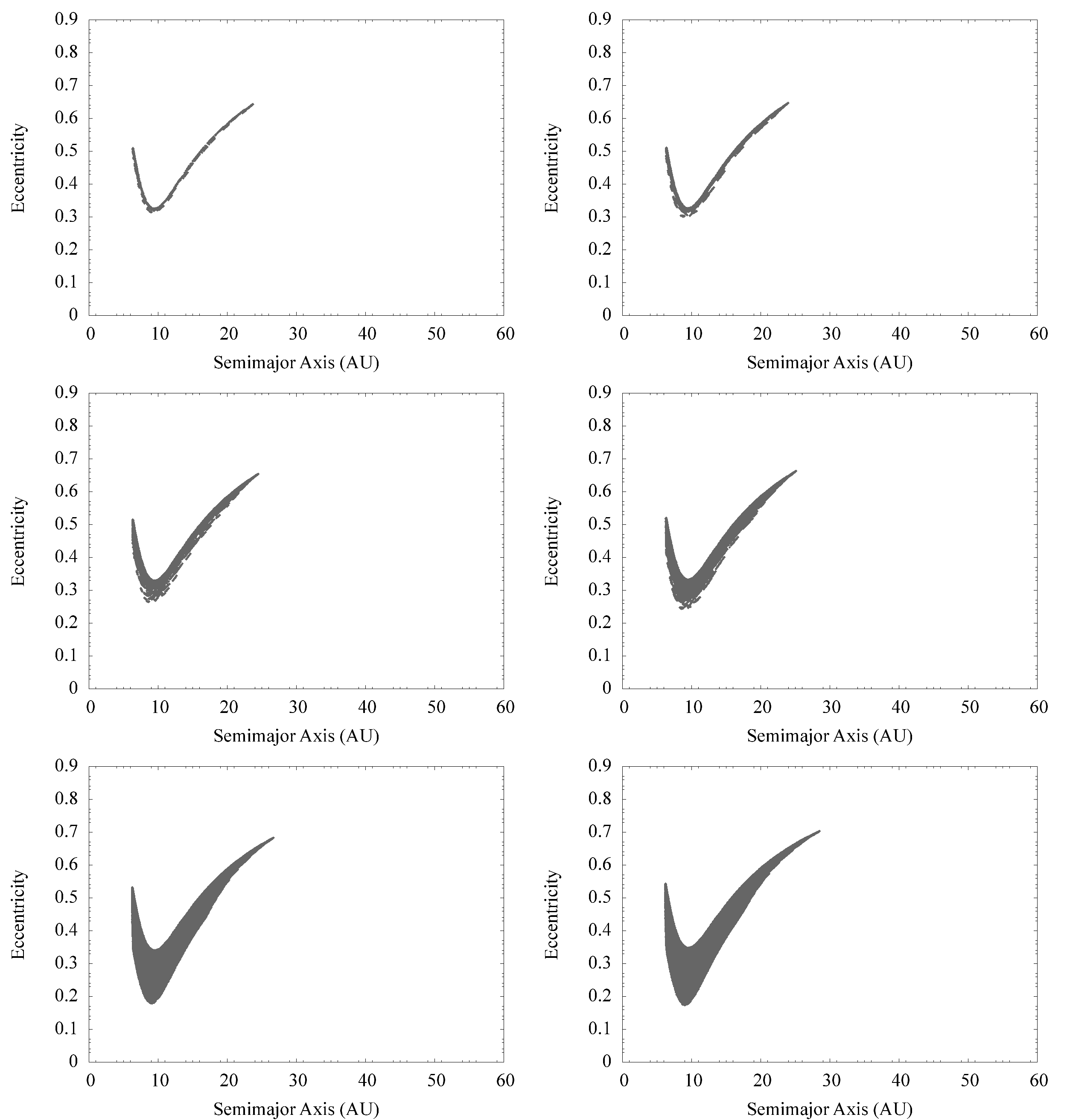}
\end{center}
\caption{Evolution of the solutions to the inverse capture problem obtained for Phoebe for increasing values of the specific impulse applied and projected in the heliocentric $a-e$ plane. From top left to bottom right, the values assumed by the specific impulse are $640,\,655,\,685,\,720,\,800,\,880$ m/s. Distances are expressed in AU.}
\label{aeplot-Phoebe}
\end{figure*}

\begin{figure*}
\begin{center}
\includegraphics[width=16cm]{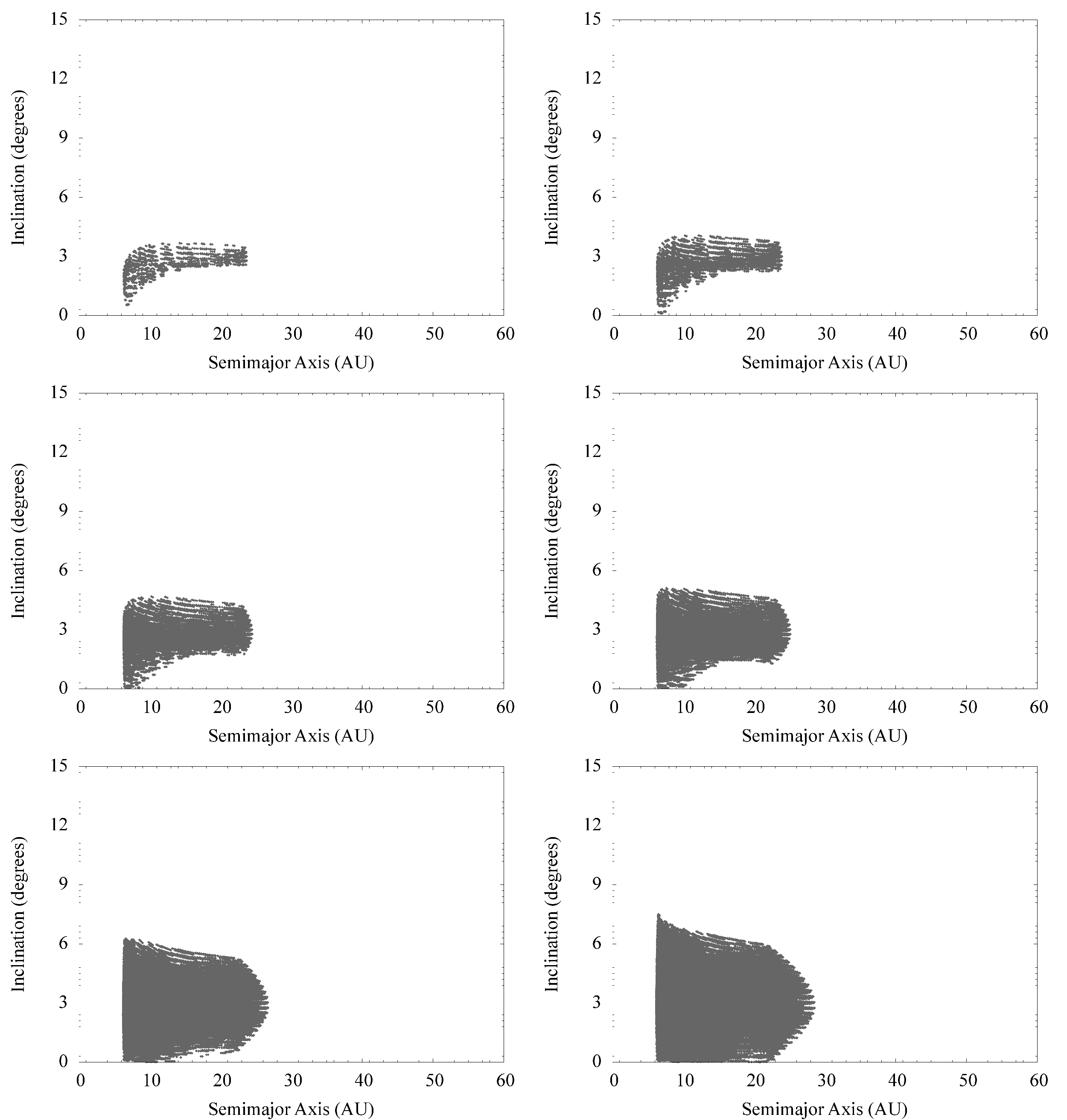}
\end{center}
\caption{Evolution of the solutions to the inverse capture problem obtained for Phoebe for increasing values of the specific impulse applied and projected in the heliocentric $a-i$ plane. From top left to bottom right, the values assumed by the specific impulse are $640,\,655,\,685,\,720,\,800,\,880$ m/s. Distances are expressed in AU while angles are expressed in degrees.}
\label{aiplot-Phoebe}
\end{figure*}

\begin{figure*}
\begin{center}
\includegraphics[width=16cm]{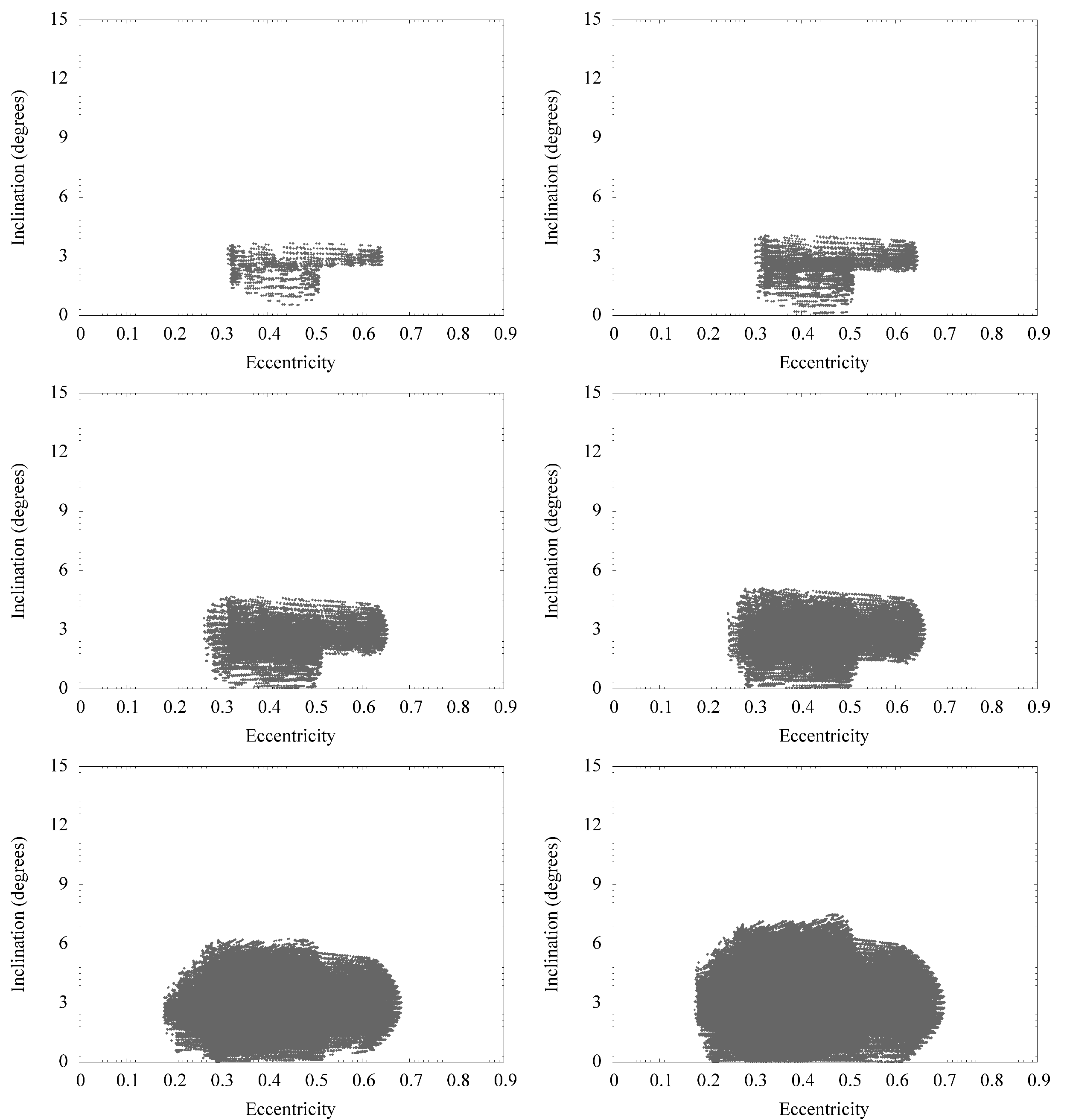}
\end{center}
\caption{Evolution of the solutions to the inverse capture problem obtained for Phoebe for increasing values of the specific impulse applied and projected in the heliocentric $e-i$ plane. From top left to bottom right, the values assumed by the specific impulse are $640,\,655,\,685,\,720,\,800,\,880$ m/s. Angles are expressed in degrees.}
\label{eiplot-Phoebe}
\end{figure*}

\begin{figure*}
\begin{center}
\includegraphics[width=16cm]{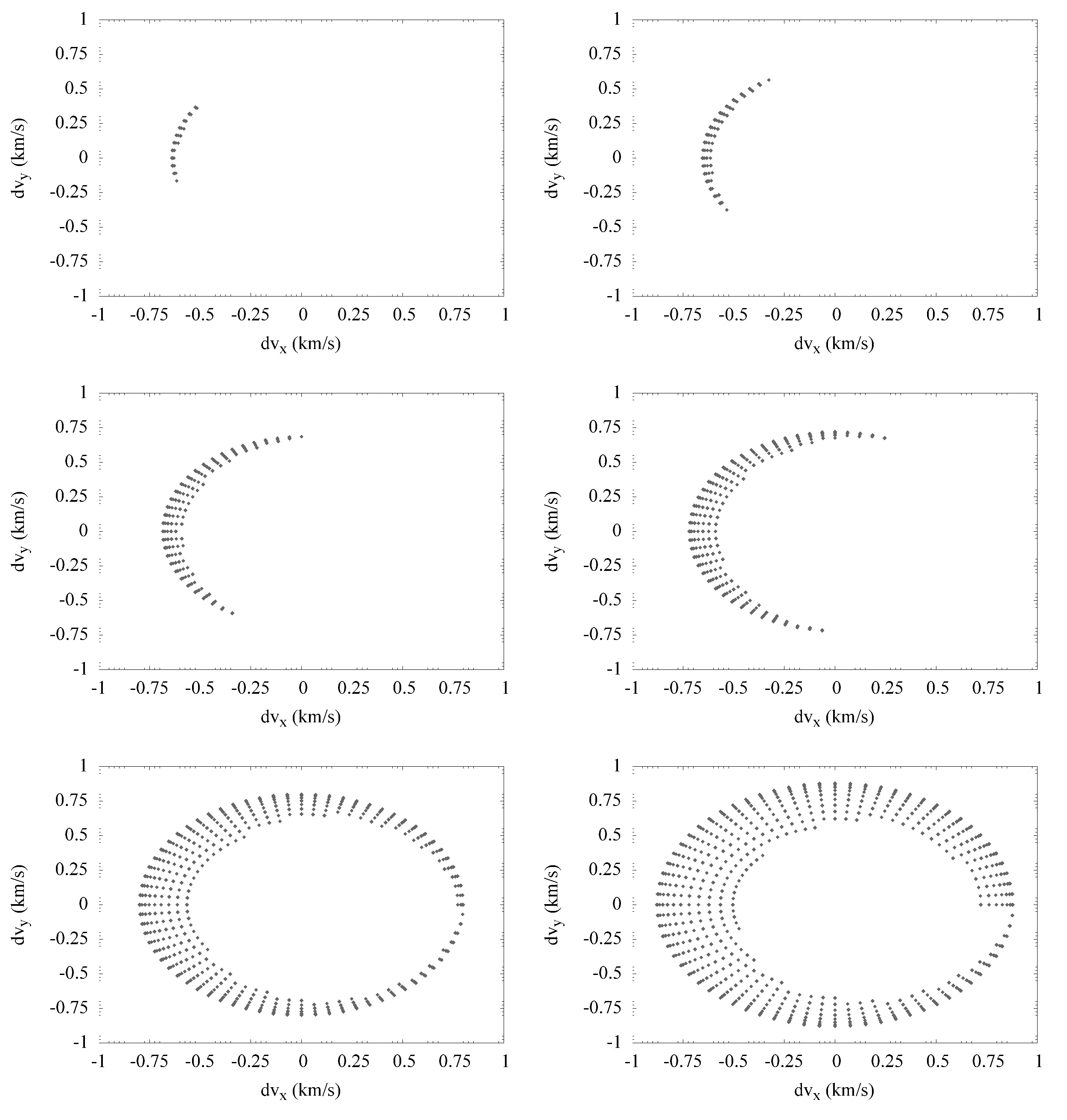}
\end{center}
\caption{Specific impulse applied to Phoebe to solve the inverse capture problem projected in the $v_{x}-v_{y}$ plane. From top left to bottom right, the values assumed by the specific impulse are $640,\,655,\,685,\,720,\,800,\,880$ m/s. Velocities in the graphs are expressed in km/s.}
\label{dvplot-Phoebe}
\end{figure*}

\begin{figure*}
\begin{center}
\includegraphics[width=16cm]{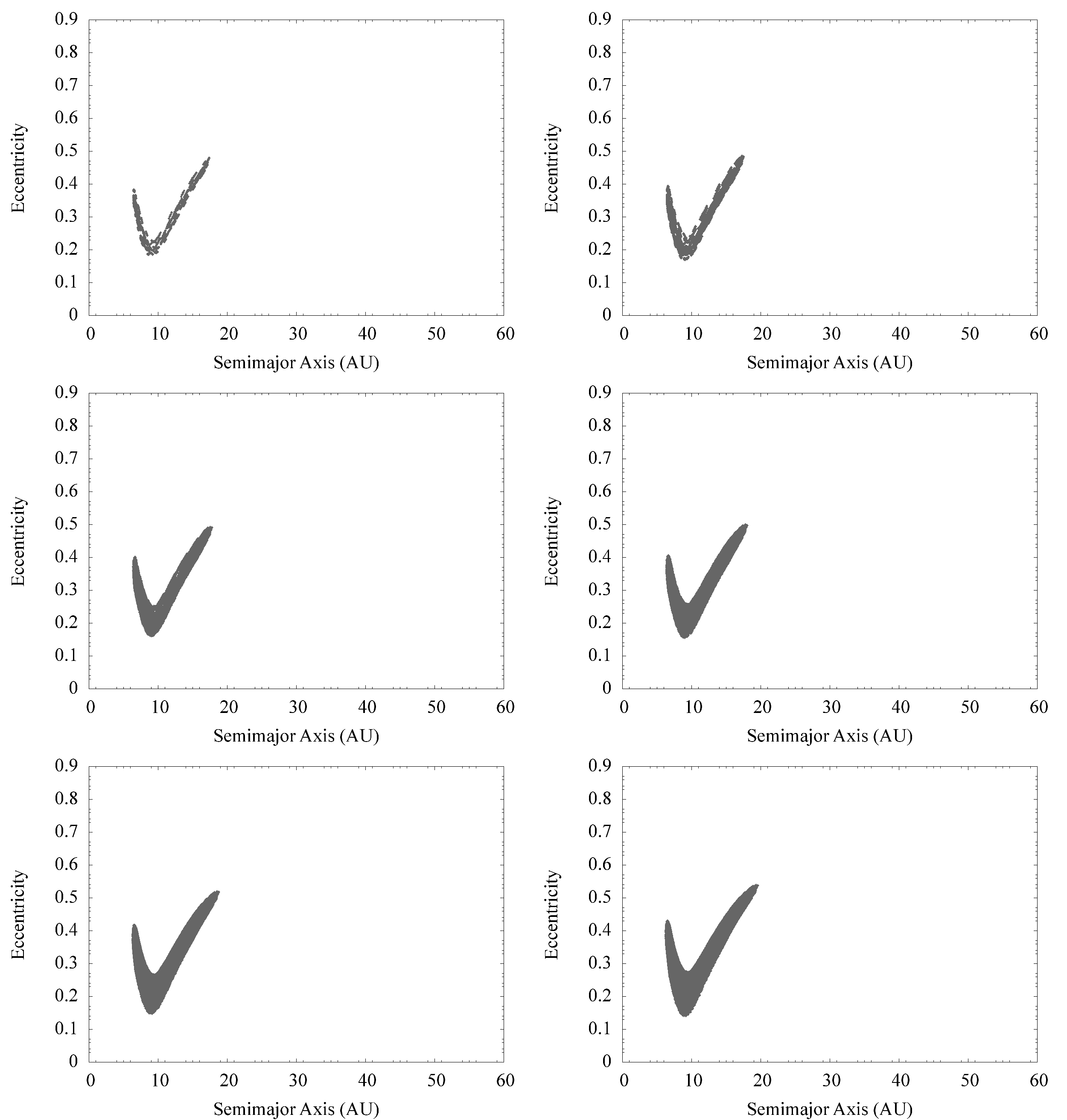}
\end{center}
\caption{Evolution of the solutions to the inverse capture problem obtained for Mundilfari for increasing values of the specific impulse applied and projected in the heliocentric $a-e$ plane. From top left to bottom right, the values assumed by the specific impulse are $515,\,530,\,560,\,590,\,670,\,750$ m/s. Distances are expressed in AU.}
\label{aeplot-Mundilfari}
\end{figure*}

\begin{figure*}
\begin{center}
\includegraphics[width=16cm]{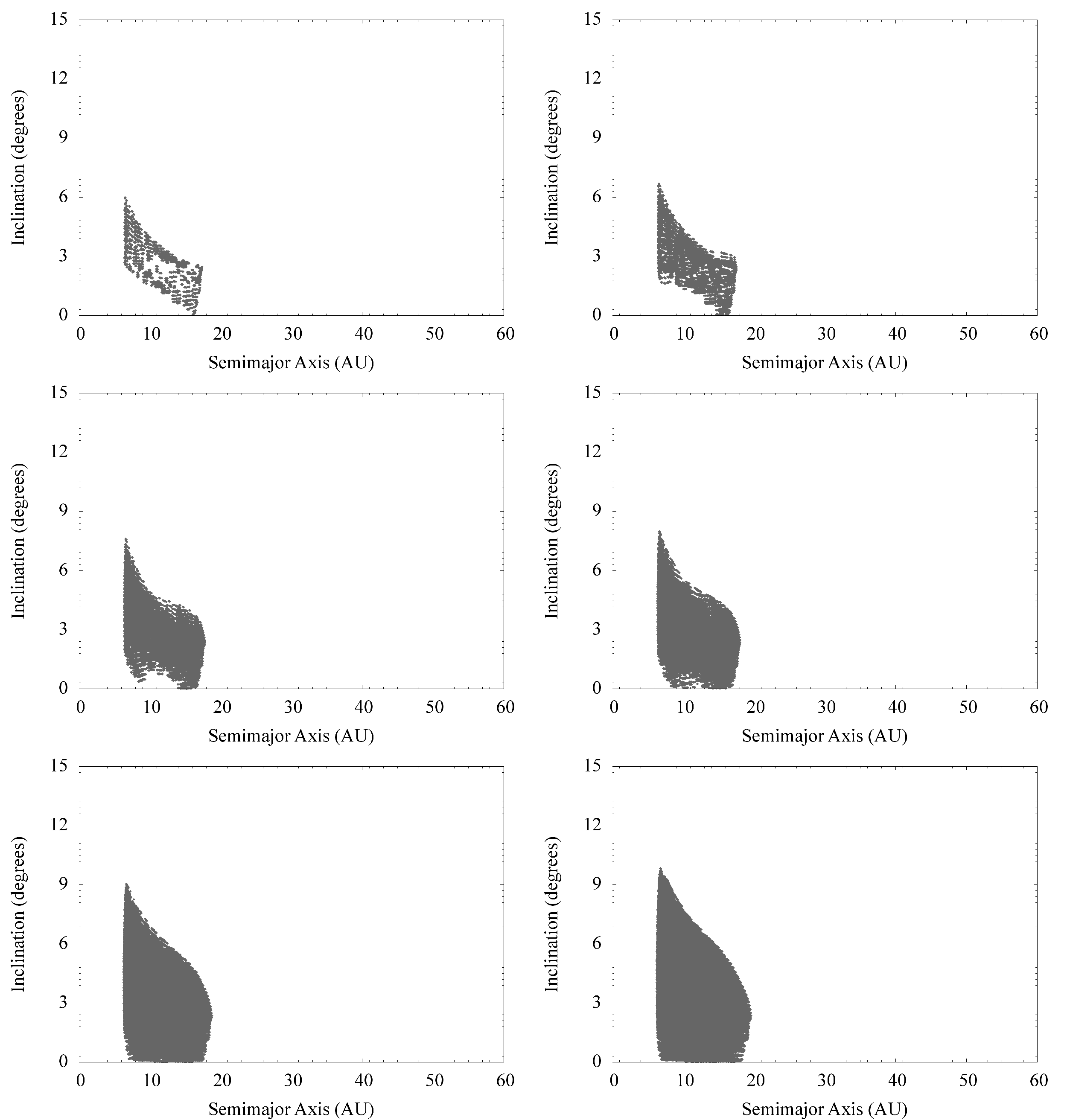}
\end{center}
\caption{Evolution of the solutions to the inverse capture problem obtained for Mundilfari for increasing values of the specific impulse applied and projected in the heliocentric $a-i$ plane. From top left to bottom right, the values assumed by the specific impulse are $515,\,530,\,560,\,590,\,670,\,750$ m/s. Distances are expressed in AU while angles are expressed in degrees.}
\label{aiplot-Mundilfari}
\end{figure*}

\begin{figure*}
\begin{center}
\includegraphics[width=16cm]{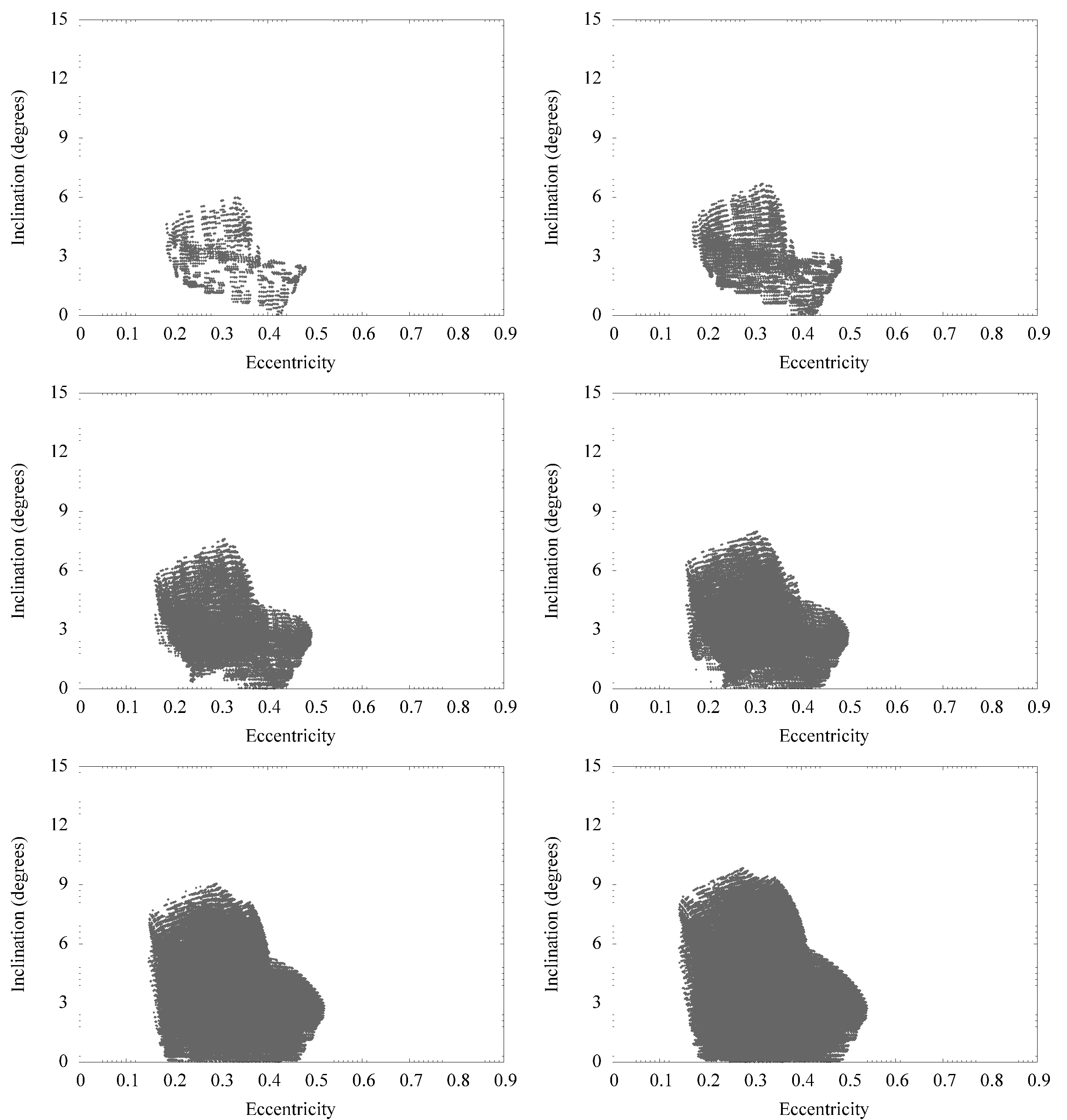}
\end{center}
\caption{Evolution of the solutions to the inverse capture problem obtained for Mundilfari for increasing values of the specific impulse applied and projected in the heliocentric $e-i$ plane. From top left to bottom right, the values assumed by the specific impulse are $515,\,530,\,560,\,590,\,670,\,750$ m/s. Angles are expressed in degrees.}
\label{eiplot-Mundilfari}
\end{figure*}

\begin{figure*}
\begin{center}
\includegraphics[width=16cm]{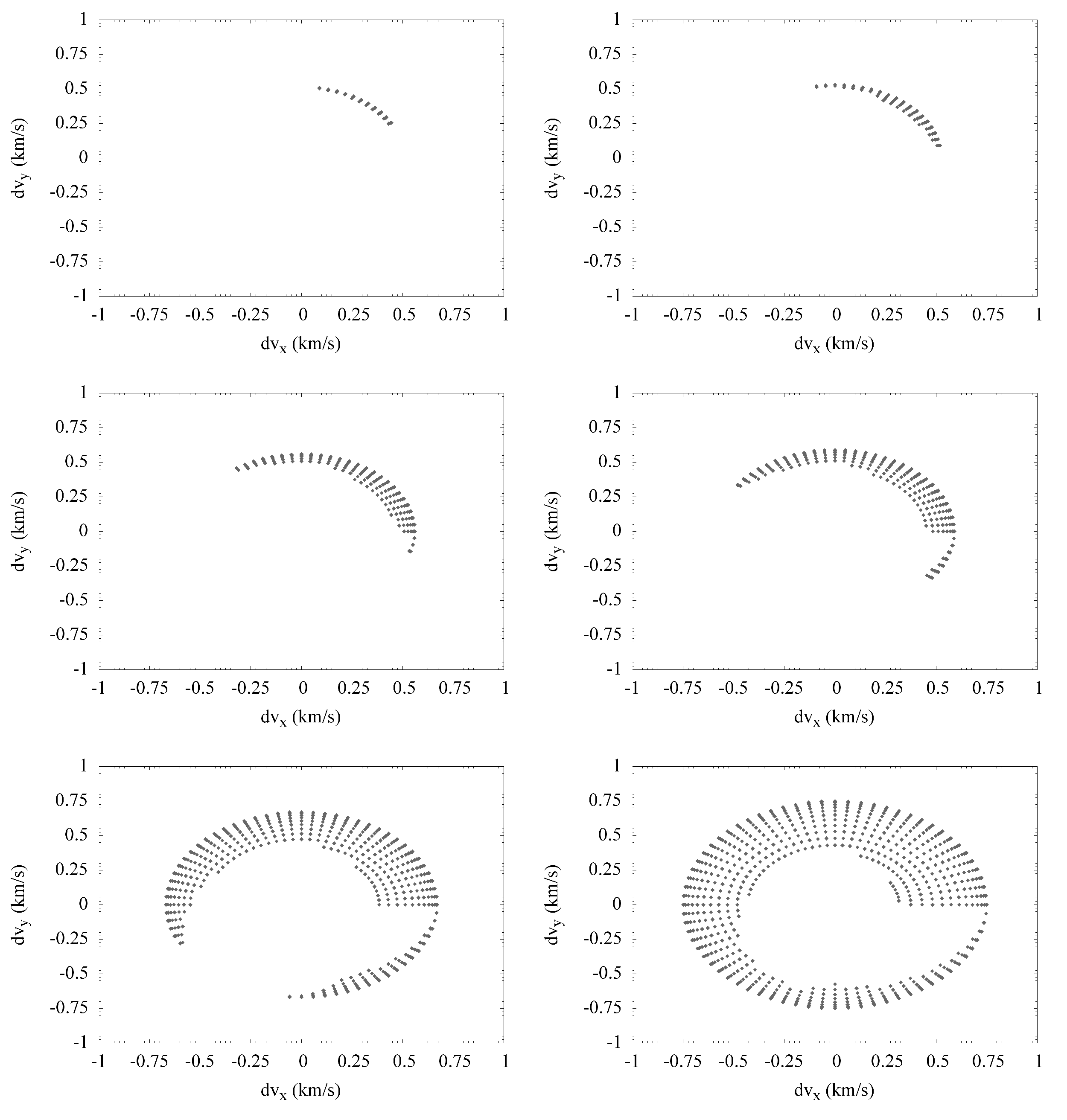}
\end{center}
\caption{Specific impulse applied to Mundilfari to solve the inverse capture problem projected in the $v_{x}-v_{y}$ plane. From top left to bottom right, the values assumed by the specific impulse are $515,\,530,\,560,\,590,\,670,\,750$ m/s. Velocities in the graphs are expressed in km/s.}
\label{dvplot-Mundilfari}
\end{figure*}

The configuration we used for our investigation of the inverse capture process of Saturn's irregular satellites has been:
\begin{itemize}
\item sampling step for the $d\vec{v_{i}}$ vectors in the $x-y$ plane: $5^{\circ}$
\item sampling step for the $d\vec{v_{i}}$ vectors in the $z$ direction: $5^{\circ}$
\item sampling step for the satellite mean anomaly: $5^{\circ}$
\item sampling step for the planetary mean anomaly: $5^{\circ}$
\item limiting value for heliocentric eccentricity: $0.9$
\end{itemize}
Since our model employed the specific impulse as its free parameter, we were not constrained by the size of the bodies in the choice of the ones to be used as our reference cases. We chose four irregular satellites (two prograde and two retrograde) representative of the main dynamical configurations for our investigation: Albiorix, Siarnaq, Phoebe and Mundilfari. As their orbits, we used the mean orbital elements we presented in Paper I. Phoebe was a natural choice as one of our case studies since our simulations showed that the secular variations of its orbit are quite regular and limited (see Paper I), implying a better preservation of its primordial dynamics. The results we obtained are presented from fig. \ref{aeplot-Albiorix} to fig. \ref{dvplot-Mundilfari}, where we showed, for increasing values of the specific impulse, the evolution of the solutions to the inverse problem in the:
\begin{itemize} 
\item heliocentric $a-e$ plane (figs. \ref{aeplot-Albiorix}, \ref{aeplot-Siarnaq}, \ref{aeplot-Phoebe}, \ref{aeplot-Mundilfari})
\item heliocentric $a-i$ plane (figs. \ref{aiplot-Albiorix}, \ref{aiplot-Siarnaq}, \ref{aiplot-Phoebe}, \ref{aiplot-Mundilfari})
\item heliocentric $e-i$ plane (figs. \ref{eiplot-Albiorix}, \ref{eiplot-Siarnaq}, \ref{eiplot-Phoebe}, \ref{eiplot-Mundilfari})
\item $d\vec{v}$ components $x-y$ plane (figs. \ref{dvplot-Albiorix}, \ref{dvplot-Siarnaq}, \ref{dvplot-Phoebe}, \ref{dvplot-Mundilfari})
\end{itemize} 
The values of the specific impulse employed for each satellite (reported in the captions of the associated figures) produced respectively $10^4,\,3\times10^4,10^5,\,2\times10^5,\,6\times10^5,\,1.2\times10^6$ solutions. These values have been chosen to represent the main features of the evolution of the solutions and range from the threshold values for which the first solutions appeared to the values for which the different families of solutions overlapped into a single continuum. We found that the minimum change in velocity to be applied in order to capture the irregular satellites lay in general between $450$ m/s and $500$ m/s, with Phoebe being a separate case requiring a value ($\approx 650$ m/s) about $30\%$ higher.\\
While the evolution of the solutions of each satellite showed its own peculiar features, there were some general conclusions we could draw. First, we noticed that the solutions were grouped into distinct and initially well separated families (see top panels of figs. \ref{aeplot-Albiorix}-\ref{dvplot-Mundilfari}) which, for increasing values of the specific impulse applied, tended to overlap to finally merge into a more or less continuous distribution (see bottom panels of figs. \ref{aeplot-Albiorix}-\ref{dvplot-Mundilfari}). Second, these families of solutions were related to preferential direction of approach, as is particularly evident from the top panels of figs. \ref{aeplot-Albiorix}, \ref{aeplot-Siarnaq}, \ref{aeplot-Phoebe} and \ref{aeplot-Mundilfari}. This occurrence is strictly related to the existence of preferential directions of impulse change (i.e. impact geometry) as is apparent from the plots of figs. \ref{dvplot-Albiorix}, \ref{dvplot-Siarnaq}, \ref{dvplot-Phoebe} and \ref{dvplot-Mundilfari}. Third, the first solutions to appear for the lowest values of specific impulse applied were characterised by high eccentricity values (generally $0.3 < e < 0.7$, except in Mundilfari's case where $0.2 < e < 0.5$) and semimajor axis spread between $5-20$ AU ($5-30$ AU in Albiorix's case), a behaviour contrary to what was generally believed to occur (e.g. the reservoir of parent bodies of the irregular satellites being located near their host planets).\\
As a rule of the thumb, prograde and retrograde satellites formed two distinct groups in terms of the solutions to the inverse capture problem. If we compare the cases with the same number of solutions, the prograde cases cover a wider region of phase space than retrograde cases. The range of inclination values covered by the former is at least one third higher than the one of the latter ($0^\circ < i < 12^\circ$ vs. $0^\circ < i < 8^\circ$, see figs. \ref{aiplot-Albiorix}, \ref{aiplot-Siarnaq}, \ref{aiplot-Phoebe}, \ref{aiplot-Mundilfari} and figs. \ref{eiplot-Albiorix}, \ref{eiplot-Siarnaq}, \ref{eiplot-Phoebe}, \ref{eiplot-Mundilfari}), the semimajor axis of the solutions reaches regions more distant from Saturn (see figs. \ref{aeplot-Albiorix}, \ref{aeplot-Siarnaq}, \ref{aeplot-Phoebe}, \ref{aeplot-Mundilfari}) and the eccentricity reaches both higher and lower values ($e < 0.2$ and $e > 0.7$ respectively, see figs. \ref{aeplot-Albiorix}, \ref{aeplot-Siarnaq}, \ref{aeplot-Phoebe}, \ref{aeplot-Mundilfari}).\\
The solutions related to the orbital region inside Saturn's orbit are the most widespread in inclination and acquire the highest inclination values (see bottom panels of figs. \ref{aiplot-Albiorix}, \ref{aiplot-Siarnaq}, \ref{aiplot-Phoebe}, \ref{aiplot-Mundilfari}).
The solutions of Albiorix, Siarnaq and Phoebe initially avoid low inclination values ($i < 3^\circ$) for those orbits coming from the outer part of the Solar System with respect to Saturn's orbit (see top panels of figs. \ref{aiplot-Albiorix}, \ref{aiplot-Siarnaq}, \ref{aiplot-Phoebe}): such inclination values are acquired for higher values of the specific impulse applied. Mundilfari had an opposite behaviour (see top panels of fig. \ref{aiplot-Mundilfari}) and avoided low inclination values for those orbits with semimajor axis located inside Saturn's orbit. There was no apparent correlation between the eccentricity and the inclination values of the solutions, with two major exceptions for those solutions associated respectively with the lowest ($e < 0.2$) and the highest (generally $e > 0.5$) eccentricity values. The former case is related to the outermost solutions, which avoid low inclination values ($i < 2^\circ-3^\circ$) and tend to cluster between $3^\circ < i < 5^\circ$ (see bottom panels of figs. \ref{eiplot-Albiorix}, \ref{eiplot-Siarnaq}, \ref{eiplot-Phoebe}, \ref{eiplot-Mundilfari}). Albiorix had a second tail in the range $5^\circ < i < 8^\circ$ (see fig. \ref{eiplot-Albiorix}, bottom panels).
The latter case is instead related to those orbits spatially located near Saturn and appears in the figures associated to the prograde satellites Albiorix and Siarnaq (see fig. \ref{eiplot-Albiorix} and fig. \ref{eiplot-Siarnaq}, bottom panels). For those solutions we observed an inverse correlation between the inclination and the eccentricity (i.e. the lower the eccentricity, the higher the inclination).
\clearpage
\clearpage
\clearpage
\clearpage
\clearpage
\clearpage
\clearpage
\clearpage
\clearpage
\clearpage
\clearpage
\clearpage
\clearpage
\clearpage
\clearpage

\section{Parent bodies of the irregular satellites}\label{comparison}

\begin{figure*}
\begin{center}
\includegraphics[width=16cm]{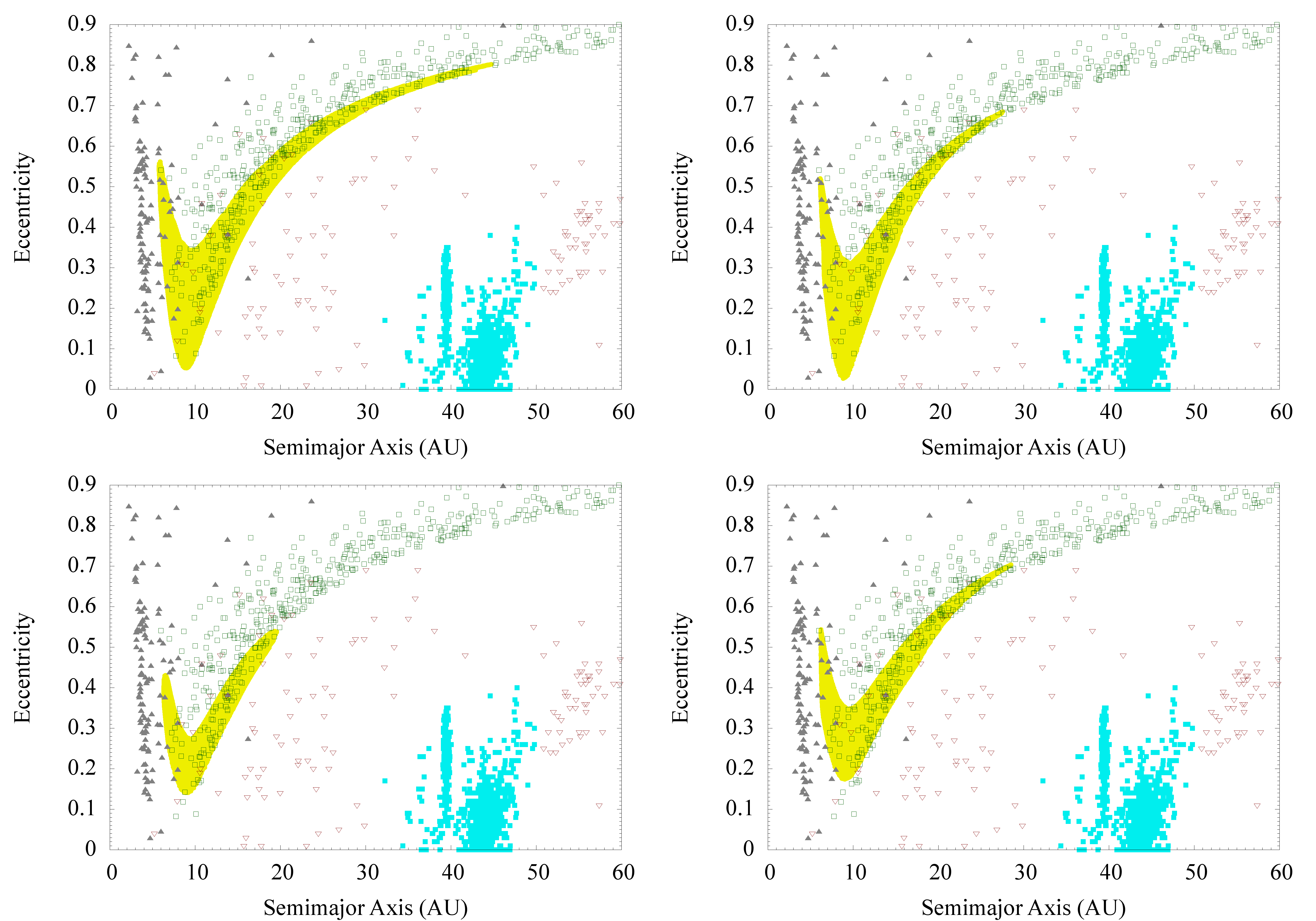}
\end{center}
\caption{Comparison in the $a-e$ plane between solutions to the inverse capture problem (yellow filled diamonds), comets (grey filled upper triangles), Centaurs and Scattered Disk objects (violet lower empty triangles), trans-neptunian objects (cyan filled squares) and Saturn Crossers (green empty squares). The irregular satellites considered were the prograde satellites Albiorix and Siarnaq and the retrograde satellites Mundilfari and Phoebe (clockwise from upper left). The specific impulse values employed for the numerical solutions of each irregular satellites are the highest ones we considered, producing $1.2\times10^6$ solutions. Distances are expressed in AU.}
\label{parentbodies-ae}
\end{figure*}

\begin{figure*}
\begin{center}
\includegraphics[width=16cm]{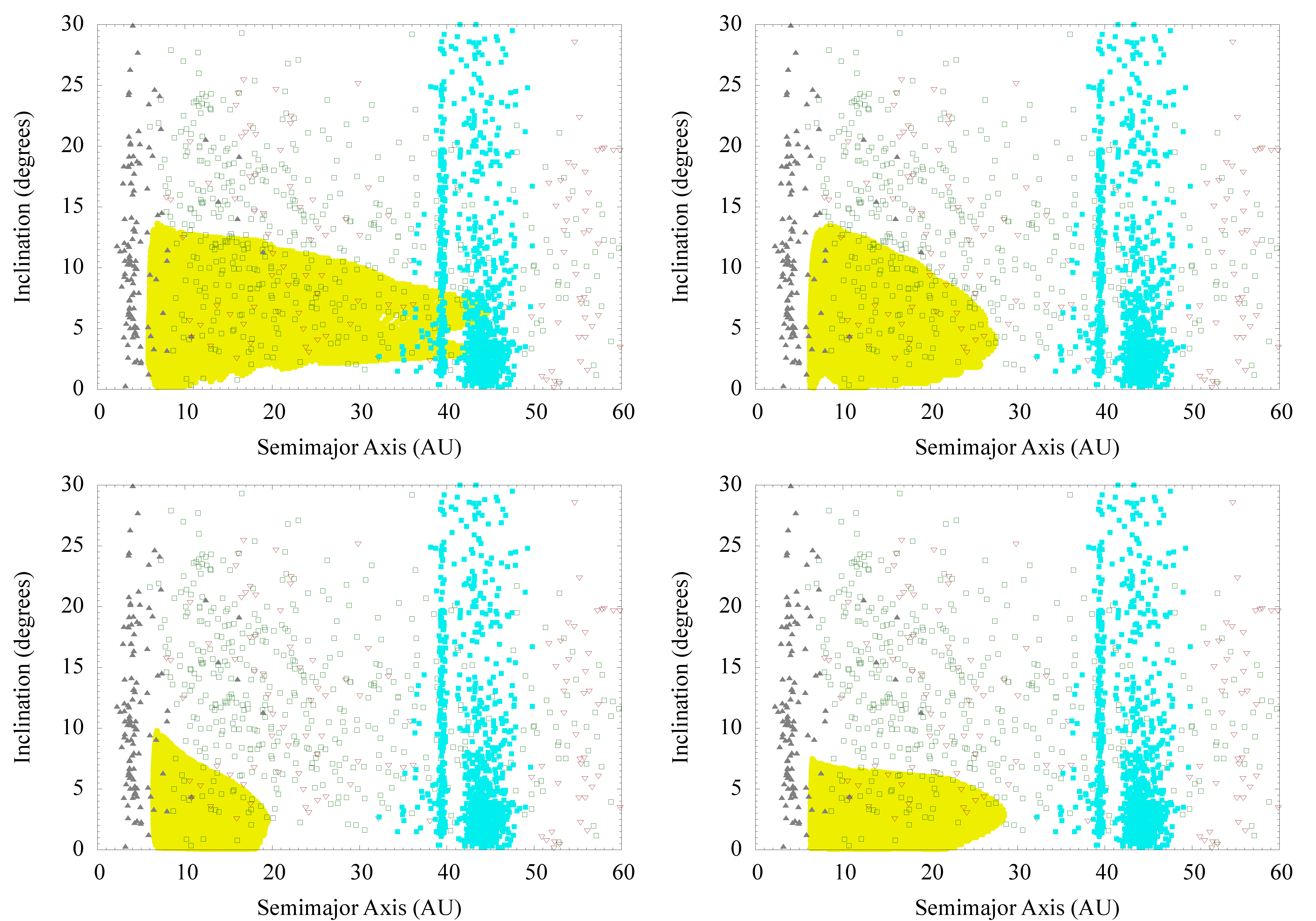}
\end{center}
\caption{Comparison in the $a-i$ plane between solutions to the inverse capture problem (yellow filled diamonds), comets (grey filled upper triangles), Centaurs and Scattered Disk objects (violet lower empty triangles), trans-neptunian objects (cyan filled squares) and Saturn Crossers (green empty squares). The irregular satellites considered were the prograde satellites Albiorix and Siarnaq and the retrograde satellites Mundilfari and Phoebe (clockwise from upper left). The specific impulse values employed for the numerical solutions of each irregular satellites are the highest ones we considered, producing $1.2\times10^6$ solutions. Distances are expressed in AU, angles are expressed in degrees.}
\label{parentbodies-ai}
\end{figure*}

\begin{figure*}
\begin{center}
\includegraphics[width=16cm]{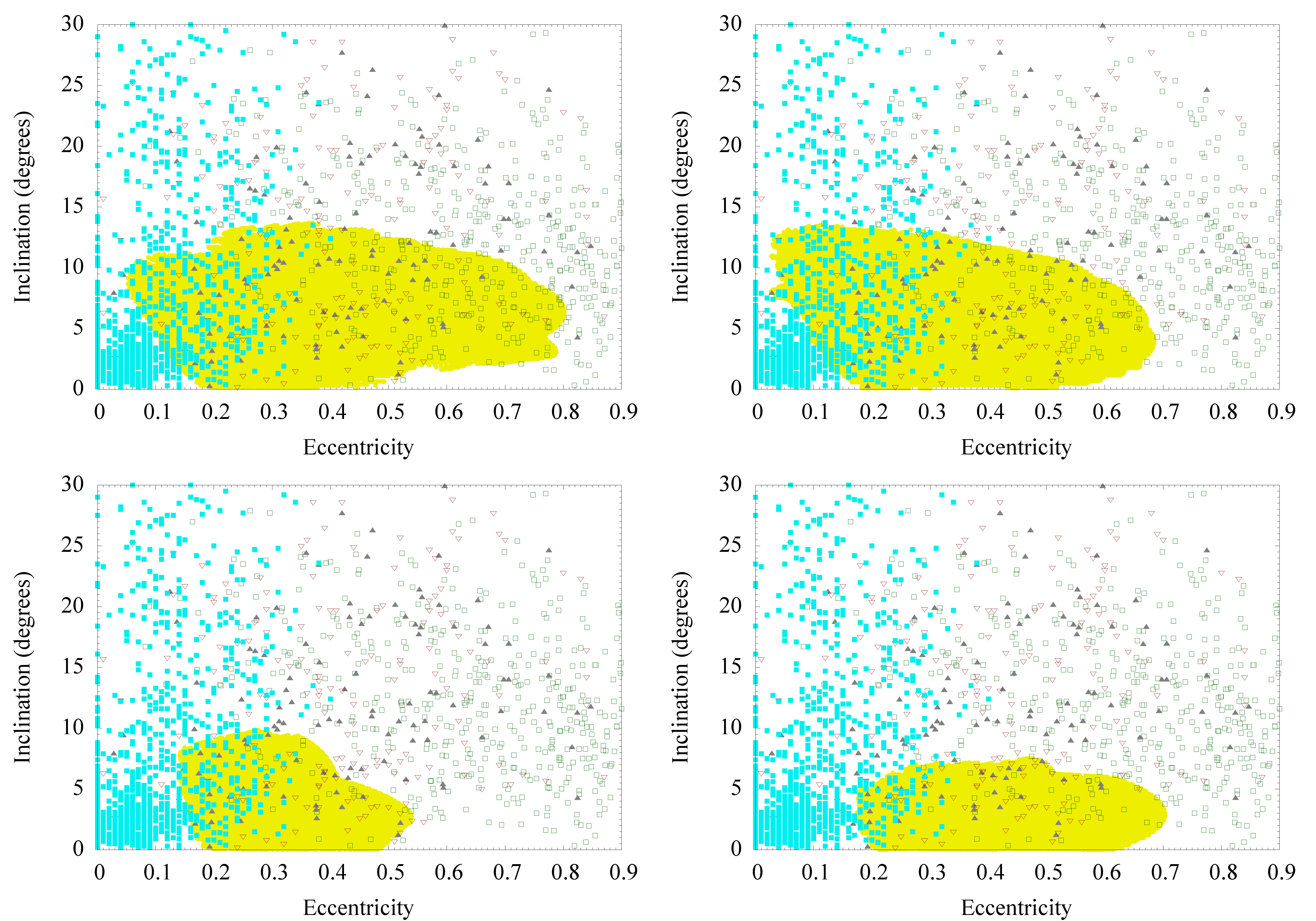}
\end{center}
\caption{Comparison in the $e-i$ plane between solutions to the inverse capture problem (yellow filled diamonds), comets (grey filled upper triangles), Centaurs and Scattered Disk objects (violet lower empty triangles), trans-neptunian objects (cyan filled squares) and Saturn Crossers (green empty squares). The irregular satellites considered were the prograde satellites Albiorix and Siarnaq and the retrograde satellites Mundilfari and Phoebe (clockwise from upper left). The specific impulse values employed for the numerical solutions of each irregular satellites are the highest ones we considered, producing $1.2\times10^6$ solutions. Angles are expressed in degrees.}
\label{parentbodies-ei}
\end{figure*}

\begin{figure}
\begin{center}
\includegraphics[width=8.4cm]{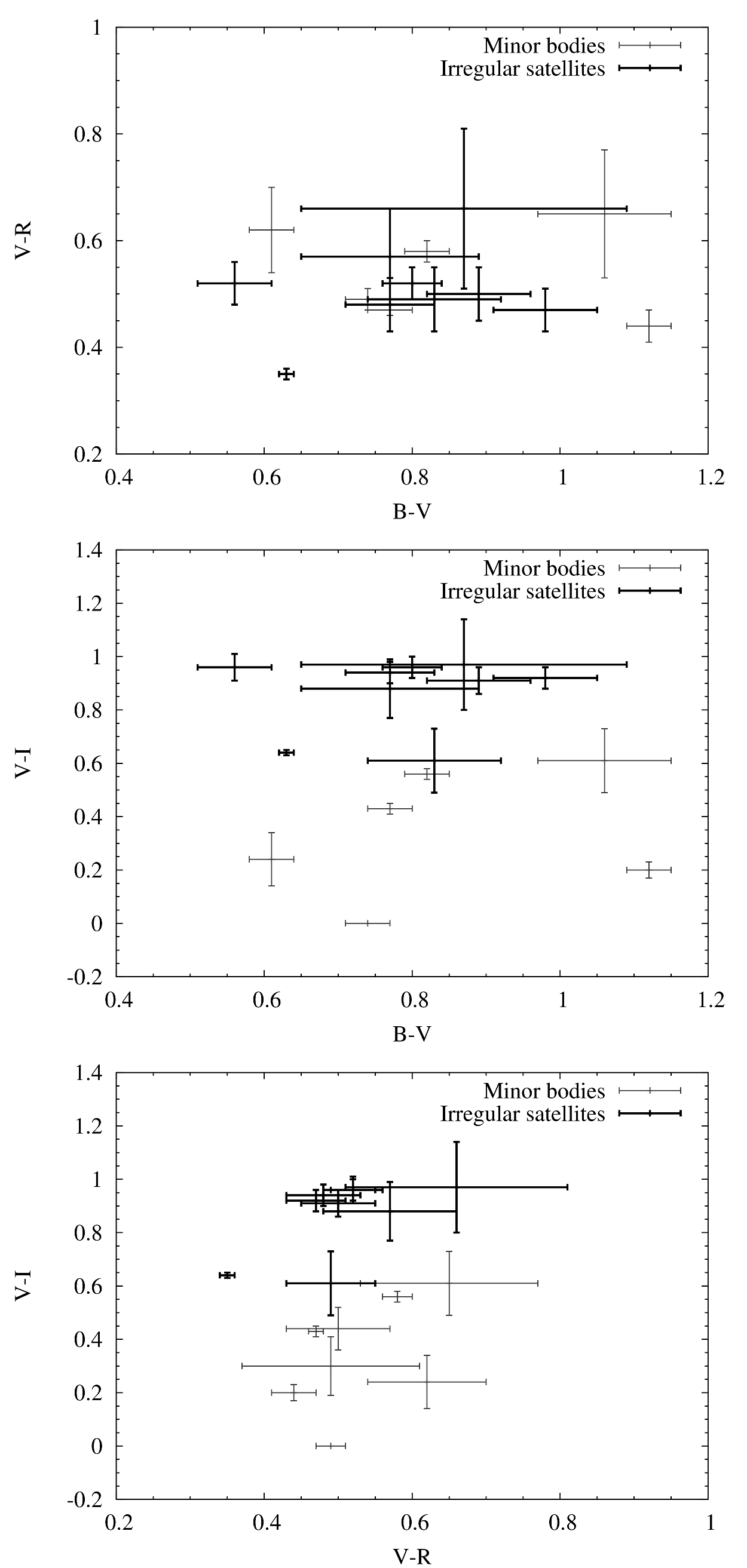}
\end{center}
\caption{Colour--colour diagrams in the $BVRI$ spectral region of irregular satellites (black symbols) and minor bodies from the catalogues considered whose orbits match the dynamical region described by the solutions to the inverse capture problem (grey symbols). The error bars represent the $1\sigma$ confidence level of the colour indexes, where available.}
\label{colour-bvri}
\end{figure}

\begin{figure}
\begin{center}
\includegraphics[width=8.4cm]{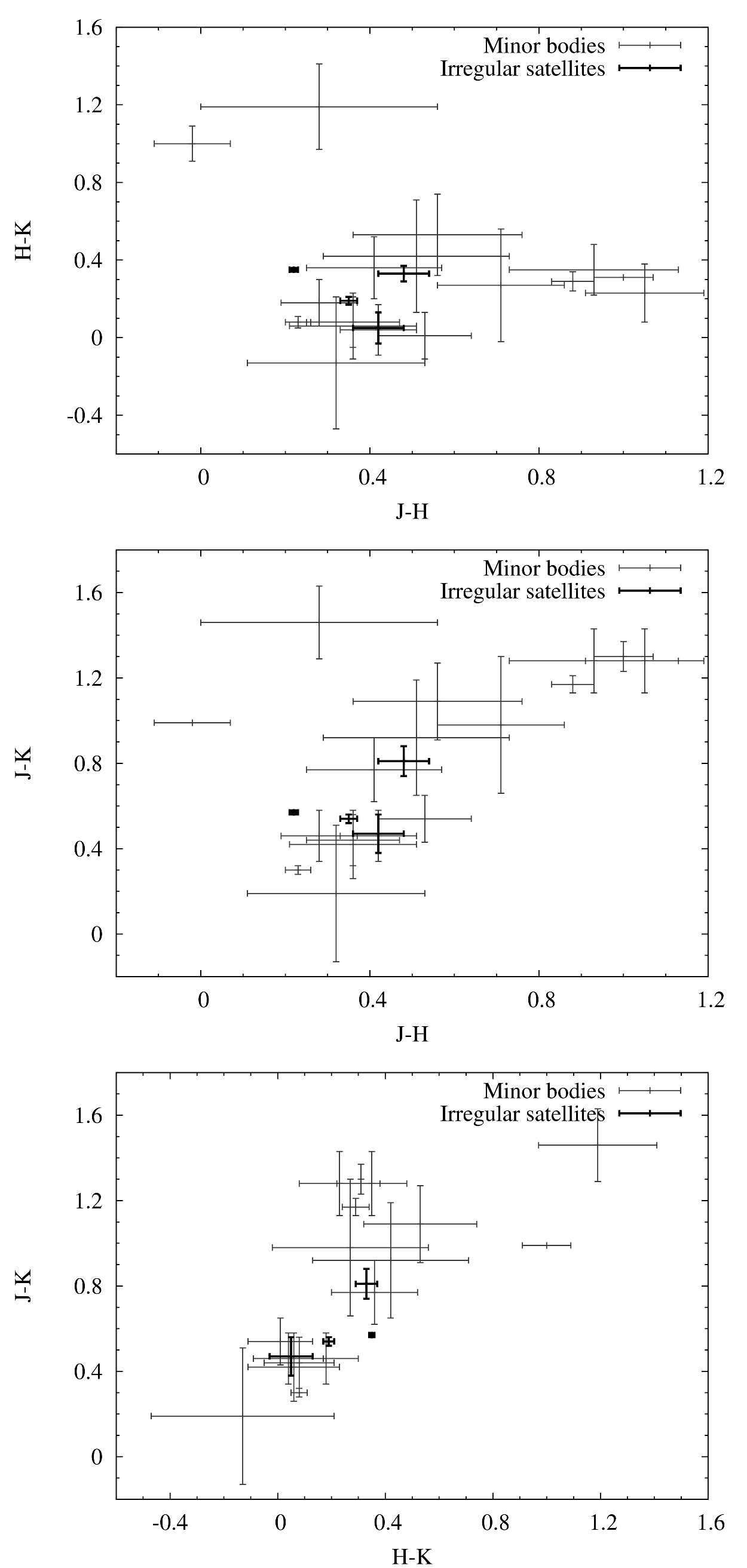}
\end{center}
\caption{Colour--colour diagrams in the $JHK$ spectral region of irregular satellites (black symbols) and comets from 2MASS catalogue whose orbits match the dynamical region described by the solutions to the inverse capture problem (grey symbols). The error bars represent the $1\sigma$ confidence level of the colour indexes, where available.}
\label{colour-jhk}
\end{figure}

Once we completed the first part of our investigation of the collisional capture problem, we proceeded to the second stage: the search for the primordial populations the parent bodies of Saturn's irregular satellites could have originated from. To reach such an ambitious goal, we compared the dynamical solutions we found and the observational data about Saturn's irregular satellites to all the available dynamical and observational data on the populations of minor bodies in the outer Solar System. Moreover, to test the capability of the collisional mechanism to explain the existence of the irregular satellites in the scenario outlined by the Nice model, we extended the comparison to the bodies crossing the orbit of Saturn in the original simulations described in \cite{gom05}, \cite{mor05} and \cite{tsi05}.
The dynamical data on the minor bodies of the outer Solar System have been obtained by the catalogues of the IAU Minor Planet Center\footnote{\url{http://cfa-www.harvard.edu/iau/}} and referred to the orbits determined at November 2006.\\
In figure \ref{parentbodies-ae}, \ref{parentbodies-ai} and \ref{parentbodies-ei} we show the solutions we found for the four Saturn's irregular satellites we used as our case study (i.e. Albiorix and Siarnaq for the prograde group and Mundilfari and Phoebe for the retrograde one). Super-imposed to the solutions we plotted the families of minor bodies of the outer Solar System: comets, trans-neptunian objects and Scattered Disk objects (respectively TNOs and SDOs) and Centaurs. We also plotted the synthetic family of Saturn Crossers found by \cite{gom05}, \cite{mor05} and \cite{tsi05} in their studies of the formation of the orbital configuration of the Solar System following a dynamical rearrangement leading to the Late Heavy Bombardment\footnote{Morbidelli, personal communication}.\\
As can be seen from the figures, three of the four families of minor bodies considered (Centaurs, comets and Saturn Crossers, TNOs and SDOs being dynamically decoupled from Saturn) overlap to various extents the region of phase space occupied by the solutions to the inverse capture problem:
\begin{itemize}
\item comets match the solutions for eccentricities in the range $0.1 < e < 0.6$ and in the radial region $6\,AU < a < 10\, AU$; the retrograde cases require $i < 8^{\circ}$, while for prograde cases inclination values $i < 12^{\circ}$ are required;
\item Centaurs partially overlap the solutions for semimajor axis $a_C$ and eccentricity $e_C$ values satisfying the relationship $a_C(1-e_C)<a_S$ with $a_S$ being Saturn's semimajor axis. The constraints on inclination are the same as found for comets;
\item Saturn Crossers partially overlap the space occupied by the solutions in the radial interval $7 AU < a < 70 AU$ for orbits having $e > 0.1$; again, the constraints on inclination are the same as found for comets.
\end{itemize}
Once we completed the dynamical comparison of the solutions to the minor bodies of interest, we proceeded in verifying if the match we found extended also to the spectrophotometric data (i.e. the colour indexes).\\
Various authors \citep{gra03,gra04,bea05} already performed ground--based photometric observations of the irregular satellites of Saturn and compared their colours with the ones of Centaurs, TNOs and recently with cometary nuclei and dead comets 
\citep{she06,sf08}. The results of such comparisons can be summarised as follows:
\begin{itemize}
\item irregular satellites present colours comparable with those of grey TNOs and of Centaurs;
\item irregular satellites lack the so--called ultrared matter characteristic of the spectra of many TNOs and Centaurs;
\item irregular satellites have colours distributions similar to those of cometary nuclei. 
\end{itemize}
Taking advantage of the results of our dynamical comparison, we devised the following strategy: instead of comparing the whole populations of Centaurs, comets and irregular satellites as previously done, we decided to compare the observational data of the irregular satellites with those of the subsets of Centaurs and comets matching the dynamical range of the solutions to the inverse problem. The aim behind our strategy was to perform a targeted comparison to minimise the effects of extraneous or of dynamically incompatible bodies, which could invalidate the global picture.\\
We proceeded by first gathering all the information available in the literature through the major and most updated catalogues on the minor bodies in the outer Solar System:
\begin{itemize}
\item MBOSS (Minor Bodies of the Outer Solar System\footnote{\url{www.sc.eso.org/~ohainaut/MBOSS/}})
\item 2MASS Comet Survey\footnote{\url{http://www.ipac.caltech.edu/2mass/}}
\item COMETS II catalogue\footnote{\cite{lam04}}
\item 2MS (Meudon Multicolour Survey\footnote{\cite{dsa06}})
\end{itemize}
The collected data were then cross--correlated with the dynamical information from the IAU Minor Planet Center's catalogues, to link each body to its orbits.\\
The observational data on Saturn's irregular satellites were taken from \cite{gra03} (\textit{BVRI} bands) and from \cite{gra04} (\textit{JHK} bands).\\
In considering the observational data we concentrated on the colour indexes instead of the spectral data because the spectra of irregular satellites and minor bodies of the outer Solar System obtained from Earth referred to too small a number of bodies and to too small a spectral range to allow for significant comparisons. Moreover, we could not take advantage of the high--resolution spectra produced by Cassini for Phoebe due to the fact that their very resolution and their wider spectral range made them unsuitable for a comparison with the range--limited, ground--based data.\\
As can be seen from fig. \ref{colour-bvri} (\textit{BVRI} bands) and fig. \ref{colour-jhk} (\textit{JHK} bands, only irregular satellites and comets from 2MASS are plotted), by considering $1\sigma$ error intervals we found a good degree of overlapping between the two groups of bodies in the $B-V$ vs $V-R$ plane and in the \textit{JHK} colours. We found a less satisfying match in the planes $B-V$ vs $V-I$ and $V-R$ vs $V-I$ due to the $V-I$ colour: however, the $V-I$ colours of irregular satellites and minor bodies are compatible at the $3\sigma$ level.\\
Such results are encouraging since they seem to validate our dynamical investigation of the inverse capture problem, but we should consider them with care: even if such a matching does exist, it exists between the present day populations of these families. While the colour indexes of the irregular satellites should have changed more or less homogeneously due to their evolving under constant environmental conditions, we cannot say the same for comets and Centaurs. Their dynamical lifetimes of the order of $10^{8}$ years mean that what we see today is a cosmogonically young population that could be not representative (at least from a compositional point of view) of the past ones. Moreover, the different orbital regions crossed during their lifetimes imply that they evolved under different environmental conditions.

\section{Comparison with Cassini data on Phoebe}\label{phoebe}

As we pointed out in the previous section, a direct comparison of the compositional data supplied by the Cassini VIMS instrument with the ones available for the other populations of minor bodies is hardly possible at the present moment because of observational issues. Nevertheless, we found that the ISS images of Phoebe could supply us with a number of informations extremely precious in the context of our research: the models of Phoebe's craters produced by the ISS images by \cite{gie06} gave us in fact the first physical evidence in support of the collisional capture scenario.\\
Without information on the dating of the cratering events, the interpretation of Phoebe's collisional history is not straightforward. We can divide the collisional capture scenarios into two main groups: catastrophic and non--catastrophic scenarios. In catastrophic scenarios, Phoebe would be the largest remnant of a bigger parent body shattered during the capture event. The craters on the surface of the satellite would be the result of the re-accretion of the smallest collisional shards produced in the impact.
The results we presented in Paper I indicate that Phoebe would be highly efficient in removing near-by objects in the region of phase space the collisional shards would populate, explaining both the intense cratering on the surface of the satellite and the absence of other remnants of the parent body. Non--catastrophic scenarios would imply that Phoebe and its parent body share a similar mass, modified only by the fragmentation and excavation effects of post-capture impacts. Both kinds of scenario suggest that the dynamical evolution of Phoebe could have suffered major changes in the past as a consequence of impact events.\\
Catastrophic scenarios are difficult to model because of the higher degeneracy in the initial conditions (e.g. the target and the impactor mass values) than non--catastrophic ones. We therefore decided to investigate the non--catastrophic scenarios and verify if the possible impacts which produced the major craters, under the right conditions, could have applied a specific impulse of the order of the ones previously estimated and thus explain the capture of the satellite. In principle, once in possession of information on the morphology, the depth and the diameter of a crater, it is possible to estimate the characteristics of the impact and the projectile by inverting the empirical formula by \cite{sch87}. Following \cite{nes03} and \cite{pem00}, we modified the formula to account for the dependence on the vertical component of impact velocity $v_{i}cos\alpha$ where $\alpha$ is the impact zenith angle \citep{gew78}:
\begin{equation}
\label{crater-volume}
V = 0.13\,\left(\frac{m_i}{\rho_t}\right)^{0.783} g^{-0.65} \left(\frac{\rho_i}{\rho_t}\right)^{0.217} (v_{i}\cos{\alpha})^{1.3}
\end{equation}
where all quantities have to be evaluated in cgs units and:
\begin{itemize}
\item $m_i,\rho_i$ are the mass and density of the impactor
\item $\rho_t$ is the density of the target
\item $v_i$ is the velocity of the impactor
\item $\alpha$ is the incidence angle measured from the zenith
\item $g$ is the surface gravity of the target
\item $V$ is the volume of the crater
\end{itemize}
In general, however, due to the degeneracy of the initial conditions it's not possible to obtain a deterministic solution to the problem. Moreover, Phoebe's case is complicated by the fact that Cassini images revealed the existence of at least $7$ major craters, distributed inhomogeneously on the surface of the satellite and with diameters greater than $50$ km (in two cases, Jason and Eurytus craters, the diameter is about $100$ km). We concentrated our attention on Jason crater, for which the resolution and the completeness of Cassini data allowed the development of a structure model \citep{gie06}. The analysis of Hylas crater ($30$ km diameter, \cite{gie06}) revealed a depth to diameter ratio $1:6$, while for the minor craters this value is about $1:5$. Since equation \ref{crater-volume} is strictly appliable only to geometrically simple, bowl--shaped craters while Jason crater shows a complex structure, made more difficult to interpret by the presence of landslides and of a secondary crater superimposed, we relied on the following approximation. We assumed a bowl--shaped structure using the diameter estimated by Cassini measures ($101$ km) and a depth to diameter ratio of $1:7$. This value was chosen to make a conservative estimate of Jason's volume and to take into account its complex morphology. The depth estimated this way ($14.43$ km) is compatible with the range of values ($13-20$ km) reported by \cite{gie06}.
Through the formula
\begin{equation}
\label{calsfer-volume}
V = \pi h \times \left( \frac{r^{2}}{2} + \frac{h^2}{6} \right)
\end{equation}
we obtained an estimated volume of $5.94\times10^{19}\,cm^{3}$, with a total excavated mass equal to $9.68\times10^{19}\,g$ if we assume its density equal to Phoebe's mean density ($1.63\,g/cm^{3}$, \cite{por05}). Since Phoebe's mass, estimated by using the mean radius ($106.6$ km, \cite{por05}) and the mean density supplied by Cassini measures, is $8.26\times10^{21}\,g$, we can easily verify that, for ejection velocities of the order of $100\,m/s$ as estimated by \cite{ben99}, the total contribution to the change in momentum due to the excavated fragments is negligible with respect to the specific impulse necessary for Phoebe's capture.\\
By inverting eq. \ref{crater-volume} we obtain a relationship linking the impact speed to the characteristics of both impact and crater:
\begin{equation}
\label{impact-velocity}
v_{i} = \left( \frac{V}{0.13\,\left(\frac{m_i}{\rho_t}\right)^{0.783} g^{-0.65} \left(\frac{\rho_i}{\rho_t}\right)^{0.217}(\cos{\alpha})^{1.3}} \right)^{0.769}
\end{equation}
In the analysis we used the value of $5\,cm/s^2$ for Phoebe's surface gravity \citep{por05} and we attributed to the density of the impactor the same value as Phoebe's mean density. The free parameters in our analysis were the mass of the impactor $m_{i}$ and the incidence angle $\alpha$. Note that the values of both densities and of the surface gravity, due to their range of variation, do not change the order of magnitude of the results.\\
Before proceeding with the analysis we want to spend a few words on the link between the impact velocity and the specific impulse necessary to capture the satellite. The impact speed can be related to the impulse transmitted to Phoebe from the collision by considering the motion of both impactor and target in a reference frame co--moving with the latter \emph{before the impact}. In this reference frame the target is initially at rest while the impactor moves at the relative speed $\vec{v}_{rel}$ which, save for gravitational focusing effects, is the same as the impact speed. The gravitational effects can be evaluated by the escape velocities of the two bodies, once their masses are known or have been fixed, and we can consider them through a corrective parameter: for Phoebe, the escape speed is of the order of $100\,m/s$ \citep{por05}.\\
In general, for real, inelastic collisions we cannot solve the equations for the conservation of the linear momentum since we don't know the amount of momentum carried away by the impactor (i.e. the amount of momentum transferred) and the general efficiency of the process, which depends on the physical characteristics of the two bodies.
As a consequence, to give a rough estimate of the plausibility of the collisional capture mechanism, we initially approximated the collision as perfectly elastic and ignored the effects of the excavation and ejection of fragments (which, for the range of mass and velocity values here considered and the typical ejection speeds, would amount to less than $0.1\%$ of the total momentum). The elastic approximation implies that we expect the impactor to ricochet and bounce from the target with a velocity higher than the escape velocity: the reason behind this choice will become clear in the following. In this scenario, by varying the free parameters and taking into account the gravitational focusing effects (which, depending on the mass values considered, could vary between $100-200\,m/s$), among the possible solutions we found that we could produce the change in velocity of $\sim650\,m/s$ necessary to capture Phoebe and excavate Jason crater through highly grazing impacts ($88^{\circ} < \alpha < 90^{\circ}$). The impactors had mass values varying between one tenth of Phoebe's mass and the mass of the irregular satellite (i.e. $0.8-8\times10^{21}\,g$) and were moving at relative speeds between $1\,km/s < \vec{v}_{rel} < 15\,km/s$ (the higher velocities relating to the lowest values of mass). For density values matching the mean density of Phoebe, that mass range would imply the size of the impactor being comprised between about $50\,km$ and $100\,km$. To improve our estimate of the plausibility of the collisional capture mechanism, we changed our approximation from perfectly elastic to inelastic (i.e. the impactor still bouncing with a velocity higher than the escape speed) by including in our equations a coefficient of restitution equal to $0.5$. We obtained the same results previously presented with a more strict requirement on the impact angle, which now varies in the range $89^{\circ} < \alpha < 90^{\circ}$.\\
The range of variations of the impact angle is at the basis of our assumptions on the degree of elasticity (i.e. elastic or inelastic, never completely inelastic) of the collisions. The results on oblique impact physics, reviewed by \cite{pem00}, show that oblique impactors ricochet retaining a significant fraction of the impact speed: if such fraction is higher than the escape velocity from the target, the impactor would bounce and escape into space \citep{pem00}. Since the lower limit of the impact velocity we found in our calculations is $1\,km/s$, an order of magnitude higher than Phoebe's escape velocity, it is highly probable that the impactor's ricochet would allow it to escape from the gravitational attraction of the irregular satellite.\\
Another important feature of oblique impacts is their lower efficiency than head-on impacts in producing high--pressure shock waves \citep{pem00}, making them less effective in disrupting the target. Taking into account the range of masses and impact velocities which could produce Phoebe's capture, this feature of oblique impacts could explain Phoebe's surviving to the capture event.\\
These results are not meant to be considered as proofs of the collisional capture of Phoebe, since the system of equations has no unique solution due to the free parameters present (i.e. the physical characteristics of the impactor and the amount of momentum carried away), but they are nevertheless the first indication of the viability of the collisional capture scenario. Moreover, we found an indirect confirmation of our results in the analysis of Jason crater performed by \cite{gie06}, who independently observed that the morphology of the crater seemed to suggest it originated following an oblique impact.\\
Before proceeding to the next and final section, we need to note that, following \cite{gew78}, oblique impacts in the range of values covered by the impact angle in both the elastic and inelastic approximations are expected to take place with a very low frequency ($n \approx 0.12\%$ in the elastic approximation and $n \approx 0.03\%$ in the inelastic one). However, the results on Phoebe's collisional efficiency in removing near-by bodies we presented in Paper I could assist us in the interpretation of this issue of the scenario: we will discuss the subject in the next section.

\section{Conclusions}\label{conclusion}

The aim of this work was to spread a new light on the origin of the system of irregular satellites of Saturn and to gain a deeper insight on the viability and the consequences of the collisional capture process.\\
To achieve these goals, we developed the MSSCC software to study the inverse capture problem and to search for the possible heliocentric pre-capture orbits. Through the use of the code, we estimated the specific impulse needed to capture the four irregular satellites we concentrated on: Albiorix and Siarnaq for the prograde group and Mundilfari and Phoebe for the retrograde one. The minimum values of the necessary impulse we found ranged between $450-500\,m/s$ with the sole exception of Phoebe which, due to its inner orbit, required a minimum value of about $650\,m/s$. We analysed the structure of the solutions in the main orbital elements space and we found that those of prograde and retrograde satellites were characterised by different features.\\
We compared the candidate primordial orbits we computed with those of the present day populations of minor bodies of the outer Solar System and with the synthetic family of the Saturn Crossers found in the original simulations of the Nice model. We found that both comets and Centaurs could supply candidates to the capture. There was also a good match between the synthetic Saturn Crossers and our solutions, thus supplying the scenario delineated by the Nice model with a natural way to populate the system of the giant planets with new satellites. We compared the observational data available on the irregular satellites with those on the minor bodies matching the dynamical range of the solutions to the inverse capture problem, and we found that, within the errors, there is enough match to strengthen our dynamical findings.\\
Finally, we used the data supplied by Cassini to test the viability of the collisional capture process in non--catastrophic impact scenarios and to constrain the impact configuration and the impactor mass by comparing our results with the information available on Jason crater. We found that highly oblique impacts could have produced Jason crater and supplied the right order of magnitude of specific impulse, the requirement on obliquity being imposed by the high specific impulse needed to capture Phoebe. The range of values of the impact angle associated with our calculations is characterised by a low probability, making a ``single collision'' scenario to appear implausible. However, in interpreting the results exposed we must take into account the following fact. Phoebe's surface indicates that the satellite had an extremely intense collisional history, with at least $7$ craters bigger than $50\,km$ and $2$ (Jason and Eurytus) of the order of $100\,km$. As a consequence, the dynamical evolution of the satellite likely suffered more than one major change in the past and the formation of Eurytus crater could be linked to Phoebe's capture as well. Since we lack detailed information on Eurytus' morphology, we cannot refine further the model. However, a ``two collisions'' scenario would relax the constraints on the impact angle, thus making the whole scenario more plausible. Moreover, the results we reported in Paper I indicate that Phoebe's efficiency in collisionally removing bodies on retrograde orbits is significant only in a limited region of phase space (i.e. the one that would be occupied by possible collisional fragments). Since we do not observe retrograde satellites in Phoebe's gap (i.e. the region comprised between $11.22\times10^6\,km$ and $14.97\times10^6\,km$ from Saturn, see Paper I) and Phoebe's sweeping effect cannot be invoked to explain this fact, the efficiency of the capture process should be low enough to explain the absence of other retrograde satellites. The low frequency associated to the needed impact parameters in the ``single collision'' scenario could help explaining why Phoebe is the only retrograde satellite residing in this orbital region.

\section*{Acknowledgements}

All the authors wish to thank the referee, Anthony R. Dobrovolskis, for help and suggestions to improve this paper. D.T. wishes to thank Alessandro Morbidelli, Torrence Johnson and Luke Dones for fruitful discussions and useful comments on the results and their interpretation.

\bsp

\label{lastpage}
\end{document}